\documentclass[aps,prb,twocolumn,superscriptaddress,showpacs]{revtex4-1}
\usepackage{epsfig}
\usepackage[dvipsnames,usenames]{color}
\usepackage{amsmath}
\usepackage{xfrac}
\emergencystretch=\maxdimen

\usepackage[colorlinks,bookmarks=false,citecolor=NavyBlue,linkcolor=OliveGreen,urlcolor=blue]{hyperref}

\usepackage{mathrsfs}
\usepackage{mathtools}
\usepackage{amsfonts}
\usepackage{amsthm}
\usepackage{amssymb}

\DeclareMathOperator{\tr}{Tr}

\newcommand{\ket}[1]{|#1\rangle}
\newcommand{\braket}[2]{\langle #1|#2\rangle}
\newcommand{\med}[1]{\left\langle #1\right\rangle}

\newcommand{\be}{\begin{equation}}
\newcommand{\ee}{\end{equation}}

\newcommand{\Z}{\mathbb{Z}}

\renewcommand{\(}{\left(}
\renewcommand{\(}{\left(}
\renewcommand{\)}{\right)}

\usepackage{bm}

\begin{document}

\title{Entanglement Hamiltonians of lattice models via the Bisognano-Wichmann theorem}

\author{G. Giudici}
\affiliation{The Abdus Salam International Centre for Theoretical Physics, strada Costiera 11, 34151 Trieste, Italy}
\affiliation{SISSA, via Bonomea 265, 34136 Trieste, Italy}
\affiliation{INFN, sezione di Trieste,  34136 Trieste, Italy}
\author{T. Mendes-Santos}
\affiliation{The Abdus Salam International Centre for Theoretical Physics, strada Costiera 11, 34151 Trieste, Italy}
\author{P. Calabrese}
\affiliation{The Abdus Salam International Centre for Theoretical Physics, strada Costiera 11, 34151 Trieste, Italy}
\affiliation{SISSA, via Bonomea 265, 34136 Trieste, Italy}
\affiliation{INFN, sezione di Trieste, 34136 Trieste, Italy}
\author{M. Dalmonte}
\affiliation{The Abdus Salam International Centre for Theoretical Physics, strada Costiera 11, 34151 Trieste, Italy}
\affiliation{SISSA, via Bonomea 265, 34136 Trieste, Italy}

\begin{abstract}
The modular (or entanglement) Hamiltonian correspondent to the half-space-bipartition of a quantum state uniquely characterizes its entanglement properties. However, in the context of lattice models, its explicit form is analytically known only for the Ising chain and certain free theories in one-dimension. In this work, we provide a throughout investigation of entanglement Hamiltonians in lattice models obtained via the Bisognano-Wichmann theorem, which provides an explicit functional form for the entanglement Hamiltonian itself in quantum field theory. Our study encompasses a variety of one- and two-dimensional models, supporting diverse quantum phases and critical points, and, most importantly, scanning several universality classes, including Ising, Potts, and Luttinger liquids. We carry out extensive numerical simulations based on the density-matrix-renormalization-group method, exact diagonalization, and quantum Monte Carlo. In particular, we compare the exact entanglement properties and correlation functions to those obtained applying the Bisognano-Wichmann theorem on the lattice. We carry out this comparison on both the eigenvalues and eigenvectors of the entanglement Hamiltonian, and expectation values of correlation functions and order parameters. Our results evidence that, as long as the low-energy description of the lattice model is well-captured by a Lorentz-invariant quantum field theory, the Bisognano-Wichmann theorem provides a qualitatively and quantitatively accurate description of the lattice entanglement Hamiltonian. The resulting framework paves the way to direct studies of entanglement properties utilizing well-established statistical mechanics methods and experiments.

\end{abstract}


\maketitle
\noindent

\section{Introduction}

Over the last two decades, entanglement has emerged as a key tool to characterize quantum phases of matter in many-body systems~\cite{vedral2008,ccd-09,ecp-10,fradkinbook,laflorencie2016}. In particular, bipartite entanglement is typically characterized by considering the reduced density matrix $\rho_A$ correspondent to a region $A$, that is obtained by tracing a state $\Psi$ over the complement of $A$ (which is denoted as $B$ in the following):
\begin{align}
\rho_A = \text{Tr}_{B}|\Psi\rangle \langle\Psi| = e^{-\tilde{H}_A}. 
\end{align}
This reduced density matrix is associated to a given entanglement (or modular) Hamiltonian (EH)~\cite{Witten:2018aa}, $\tilde{H}_A$, which shares its same eigenvectors $|\phi_\alpha\rangle$, and whose spectrum is bounded from below~\footnote{In this work, we will refer to $\tilde{H}_A$ as modular Hamiltonian if intended in the context of quantum field theory, and as entanglement Hamiltonian if intended in the context of statistical mechanics and condensed matter systems.}. The spectral properties of the EH uniquely determine the entanglement properties of the partition $A$ of $\Psi$: for instance, its spectrum - the entanglement spectrum - determines the von Neumann entropy. 

A direct knowledge of the functional form of $\tilde{H}_A$ is of tremendous utility for two main reasons. From the experimental side, it allows to measure entanglement properties of a given state via direct engineering of the EH~\cite{Dalmonte:2017aa}, in particular, in cases where direct access to the wave function is not scalable (such as in experiments requiring full state tomography) or not possible at all. Hence it provides a feasible route for the measurements of, e.g., entanglement spectra, which are experimentally challenging to access in a scalable manner~\cite{pichler2016measurement}. From the theoretical side, it immediately opens up a new toolbox to investigate entanglement properties of lattice models using conventional statistical mechanics techniques, both numerical and analytical. However, in the context of many-body lattice models, it has proven challenging to determine $\tilde{H}_A$ analytically even for free theories - the only results being the EH of the Ising chain away from criticality~\cite{Peschel_1999,peschel2009}, of some one-dimensional free fermion systems~\cite{Eisler_2017,eisler2018}, and of few other less generic examples~\cite{Peschel_1999,Nienhuis_2009}.

In this work, we provide a throughout investigation of the entanglement Hamiltonian correspondent to the ground state of lattice models based on the application of field theoretical results to microscopic theories. In particular, following Ref.~\onlinecite{Dalmonte:2017aa}, we recast on the lattice
the Bisognano-Wichmann (BW) theorem~\cite{bisognano1975duality,bisognano1976duality,guido2011modular,Witten:2018aa} and its extensions~\cite{longo1982,Brunetti1993,Casini:2011aa,Cardy:2016aa} to conformal field theory (CFT). We verify their predictive power by systematically comparing several properties of the corresponding EHs to the original lattice model results. The main result of our analysis is that this approach returns a closed-form expression for the lattice EH which accurately reproduces not only the entanglement spectrum, but also properties directly tied to the eigenvectors of the reduced density matrix, such as correlation functions and order parameters.

We carry out our analysis by combining a series of numerical approaches, including exact diagonalization, Density Matrix Renormalization Group (DMRG)\cite{White1992,schollwock2011density}, and quantum Monte Carlo simulations. At the methodological level, applying these approaches directly at the level of the BW \emph{entanglement field theory} - that is, the field theory obtained by applying the BW theorem to the lattice problem - is in principle straightforward, apart from few technical details due to the specific shape of the EH that we discuss in some detail. We focus on interacting one- (1D) and two-dimensional (2D) lattice models, spanning both quantum critical phases and points, and ordered, disordered, and symmetry-protected topological phases whose low-energy physics is captured by a quantum field theory with emergent Lorentz invariance (in the critical cases, with dynamical critical exponent $z=1$). Overall, our results support the fact that the applicability of this approach solely relies on universal properties, in particular, on how accurately the low-energy properties of a lattice model are captured by a Lorentz-invariant quantum field theory. 
Along reporting our results, we provide a comparison with model specific methods employed so far to grasp salient features (and, in some cases, the exact form) of the EH based on perturbation theory, exact solution of free fermionic problems, and perturbed CFT  techniques. We anticipate that, whenever a comparison can be drawn, previous results are in quantitative agreement with the lattice BW approach.

The structure of the paper is as follows. In Sec.~\ref{sec_bw}, we review the BW theorem original formulation, its extensions in the context of conformal field theories, and present in detail its adaption to lattice problems.  We present a qualitative discussion of the applicability regimes of this adaption, and then discuss the specific diagnostics we employ to compare the original EH result with the BW EH on the lattice, and our numerical approaches. In Sec.~\ref{sec_1d}, we discuss our results in the context of 1D systems, starting with models endowed with discrete symmetries (Ising, Potts), and then moving to spin chains with continuous symmetries (XXZ and $J_1-J_2$ models). In Sec.~\ref{sec_2d}, we focus on 2D systems, discussing in detail the Heisenberg and XY models on both cylinder and torus geometries. Finally, in Sec.~\ref{sec_concl}, we draw our conclusions, compare with alternative approaches, and point out some  perspectives and questions motivated by our approach and results. 

\section{Entanglement Hamiltonians: from field theories to lattice models}
\label{sec_bw}

In this section, we provide some background material on the BW theorem, and its adaption to lattice models. We present a general discussion on the applicability regimes of the latter approach, and describe the main criteria used for numerical checks carried out in the subsequent sections. 

\subsection{The Bisognano-Wichmann theorem and its conformal extensions}

In an arbitrary {\it relativistic quantum field theory}~\footnote{We consider field theories whose Hilbert space is in tensor product form with respect to spatial partitions. The case of gauge theories with non-trivial centre will be studied elsewhere.}, the general structure of the reduced density matrix of the vacuum state can be obtained for the special 
case of a bipartition between two half-spaces of an infinite system (i.e. $\vec{x}\equiv (x_1,x_2,\dots x_d)\in {\mathbb R}^d$ and $A=\{\vec{x}| x_1>0\}$).
The specific form of the modular Hamiltonian is given in a series of papers by Bisognano and Wichmann, which can be recast in a single, general result that we refer to as Bisognano-Wichmann theorem~\cite{bisognano1975duality,bisognano1976duality}. 
This theorem states that, for a given a Hamiltonian density $H(\vec{x})$ and a for the half-bipartition above, the modular Hamiltonian of the vacuum (ground state) is
\begin{align}
\tilde{H}_A = 2 \pi \int_{\vec{x} \in A} d\vec{x}\left(x_1 H(\vec{x}) \right) + c^{\prime},
\label{BWtheorem}
\end{align}
where $c^{\prime}$ is a constant to guarantee unit trace of the density matrix, and the speed of light has been set to unity. A first key feature of this result is that its applicability does not rely on any knowledge of the ground state, and thus can be applied in both gapped and gapless phases, and quantum critical points. A second feature is that the results is applicable in any dimensionality: this will turn particularly important below, as very little is known about entanglement Hamiltonians of lattice models beyond one-dimension. Moreover, Eq.\,\eqref{BWtheorem} has a clear-cut physical interpretation in terms
of entanglement temperature~\cite{haag2012local,Wong:2013aa,Arias:2017aa}: if we interpret $\rho_A$ as
thermal state, this corresponds to a state of the original
Hamiltonian $H$ with respect to a locally varying temperature, very large close to the boundary of $A$, and
decreasing as $1/x_1$å far from it.

In the presence of conformal invariance, it is possible to further extend the BW results to other geometries~\cite{longo1982,Brunetti1993,Casini:2011aa,Cardy:2016aa,Casini_2009}. In any dimension, it is possible to derive the modular Hamiltonian of a hyper-sphere of radius $R$.~\cite{Casini:2011aa} Here, we will be interested in three specific cases in one spatial dimension, whose EHs were obtained in Ref.~\onlinecite{Cardy:2016aa}. 
The first one concerns a finite partition of size $\ell$ embedded in the infinite line when $\tilde{H}_A$ reads\cite{Casini:2011aa,Cardy:2016aa}
\begin{align}
\tilde{H}_A^{\texttt{(CFT1)}} = 2 \pi \int_{0}^{\ell} dx\left[x\left(\frac{\ell - x}{\ell} \right) H(x) \right] + c^{\prime}.
\label{BWtheoremCFT}
\end{align}
This formula can be generalized to the case of a finite partition of length $\ell$ in a ring of circumference $L$:\cite{Cardy:2016aa}
 \begin{align}
\tilde{H}_A^{\texttt{(CFT2)}} = 2 L \int_{0}^{\ell} dx\bigg[\frac{\sin\left(\frac{\pi (\ell-x)}{L}\right)\sin\left(\frac{\pi x}{L}\right)}{\sin(\pi \ell/L)} H(x) \bigg] + c^{\prime}.
\label{BWtheoremCFT2}
\end{align}
In addition, for a finite open system of length $L$ and for a finite partition of length $L/2$ at its edge (i.e. $A=[0,L/2]$ and $B=[-L/2,0]$)  we have \cite{Cardy:2016aa}
\begin{align}
\tilde{H}_A^{\texttt{(CFT3)}} = 2 L \int_{0}^{L/2} dx\sin\(\frac{\pi x}{L}\) H(x) + c^{\prime}.
\label{BWtheoremCFT3}
\end{align}
We mention that, in the vicinity of a conformal invariant critical point, an alternative description of the EH with respect to the original BW EH has been suggested~\cite{Cho:2017aa}.

Before turning to lattice models, it is worth to stress three properties of these results. The first one is that, in its original formulation, the BW theorem relies on the existence of a cyclic vector in the field theory itself. In the case of a half-bipartition, this is not a problem on the lattice, but might become so for other, un-equally sized partitions - which we do not treat in the following. The second is that, even if the modular operator is defined only from the ground state wave-function, it contains information about the entire operator spectrum of the theory. This suggests that  universal properties of the lattice models might be encoded in the deviations (including finite size ones) of the entanglement spectra evaluated from the Lattice Bisognano-Wichmann (LBW) entanglement Hamiltonian (which we describe in the next subsection) from the exact one. The third is that the BW result implies that the EH is local, and contains only few-body terms which are already present in the original model. This fact has some immediate consequences: {\it i)} it makes a direct experimental realization of the LBW EH feasible in synthetic quantum matter setups~\cite{Dalmonte:2017aa}, and {\it ii)} it makes its direct study amenable to the same tools of statistical mechanics applicable to the original problem, at least in principle. 

\subsection{Entanglement Hamiltonians for lattice models via the Bisognano-Wichmann theorem}

Differently from the field theory case, much less is known about the entanglement Hamiltonian of ground states of lattice models. In some specific cases, direct insights can be gathered by the explicit structure of the ground state wave function. Examples include the determination of the ES and EH structure in strongly gapped phases~\cite{Calabrese_2010,Alba:2012aa,alet2014,Alba_2012}, where perturbative arguments are applicable, the EH obtained via variational wave-functions~\cite{Cirac_2011}, or the Li and Haldane argument on the structure of the ES of topological phases~\cite{Li2008,Regnault:2015aa} - which can also be understood using the BW theorem~\cite{Swingle2012}. Similar arguments can be applied to wave functions with very short correlation length $\xi$, as in those cases, the EH becomes essentially a projector for distances beyond $\xi$. Other fundamental insights could come from the related concepts of entanglement contour~\cite{Botero_2004,Chen_2014,Fr_rot_2015,Coser_2017,Tonni_2018}, probability distribution of the entanglement spectrum~\cite{Calabrese_2008,pm-10,Alba_2017}, and relative entropy~\cite{Blanco2013,Lashkari:2014aa,Balasubramanian2015,Jafferis2016,Ruggiero2017}. 

\begin{figure}[]
{\centering\resizebox*{8.7cm}{!}{\includegraphics*{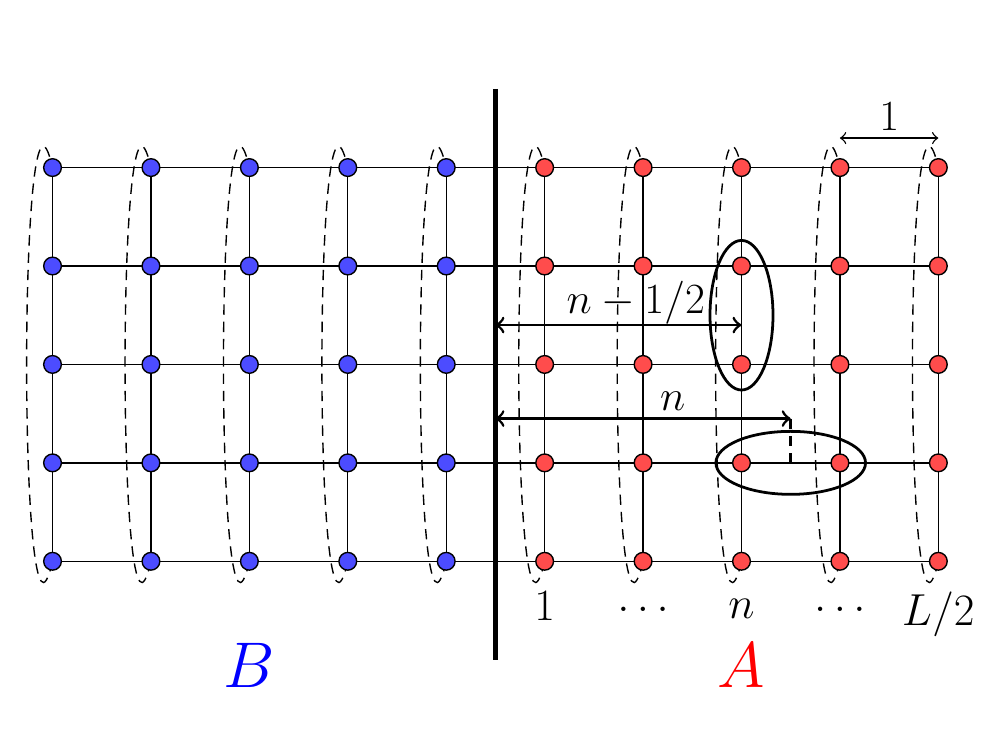}}}
\caption{Sketch of the lattice configuration for two-dimensional systems: we consider systems with periodic boundary conditions along the $y$ (vertical) direction, and either open or periodic boundary conditions along the $x$ (horizontal) direction, of length $L$. The system bipartitions we consider are defined by $A=\{(x,y)|x\in[1, L/2]\}$. The distance from the boundary (Eq.\,\eqref{eq_dist}) corresponding to different Hamiltonian terms (indicated by encircled pairs) is portrayed schematically as the geometric distance of the center of the bond from the boundary.} 
\label{sec2fig1} 
\end{figure}

Exact results without assuming any structure of the ground state wave-function have been derived only in few 1D free theories \cite{Peschel_1999,peschel2009,Eisler_2017,eisler2018}. As we discuss below, these results are very suggestive about the correctness (and, at the same time, indicate potential limitations) of the LBW EH we will discuss in the next subsections.  

Our goal here is to provide a generic recipe to derive an approximate but very accurate (in particular, able to capture all universal features) EH of a lattice model without specific knowledge of the ground state wave-function. As the starting point, following Ref.~\onlinecite{Dalmonte:2017aa}, we recast the BW theorem and its conformal extension on the lattice, formulating simple candidate EHs. 
Explicitly, let us consider a lattice model in one or two dimensions with on-site and nearest-neighbour couplings:
\begin{align}
 H = & \Gamma\sum_{x,y,\delta=\pm1} \left[ h_{(x,y),(x+\delta,y)}  +  h_{(x,y),(x,y+\delta)} \right] \nonumber\\
 + &\Theta \sum_{x,y} l_{(x,y)},
 \label{Ham}
\end{align}
where $\Gamma$ is a homogeneous  coupling (e.g., exchange term) and $\Theta$ is an on-site term (e.g., transverse or longitudinal field).
The spatial coordinates are defined as $x,y\in\{-L/2+1,\dots L/2\}$, where $L$ is the linear size of the system, which we fix to be even.
For one-dimensional systems (read just the $x$ coordinate in the aforementioned expression) we study systems with both open (OBC) and periodic boundary conditions (PBC),
while for two-dimensional systems we consider finite cylinder and torus geometries, see Fig.\,\ref{sec2fig1}.

Let us now split the system into two equal halves;
the corresponding lattice Bisognano-Wichmann EH (LBW-EH) is then given by
\begin{align}
 \tilde{H}_{A,BW} = \beta_{EH}  &\sum_{x,y,\delta=\pm1} \left(\Gamma_x h_{(x,y),(x+\delta,y)}  + \Gamma_y h_{(x,y),(x,y+\delta)} \right) \nonumber \\
              +  &\sum_{x,y} \Theta_{x,y} l_{(x,y)} ,
 \label{BW-EH}
\end{align}
where the inhomogeneous couplings and on-site terms depend on the distance from the  boundary separating subsystem $A$ and $B$ (see Fig.\,\ref{sec2fig1}) 
according to the geometry of the original system.
In the case of a 1D with OBC or for the cylinder geometry in 2D, the BW theorem in Eq.~\eqref{BWtheorem} suggests 
\begin{eqnarray}\label{eq_dist}
\Gamma_x &=& x\Gamma, \nonumber  \\
\Gamma_y &=& \left(x - \frac{1}{2}\right) \Gamma, \nonumber \\
\Theta_{(x,y)}& =& \left(x - \frac{1}{2}\right) \Theta. 
\end{eqnarray}
This putative EH is expected to provide extremely accurate results being just the lattice discretization of Eq.~\eqref{BWtheorem}, 
at least in the limit $L/2\gg\xi$ when finite size effects should be be negligible. 
However, in the following, we will use this EH also for some critical cases in order to check how this copes with finite volume effects: this is a fundamental exercise in view of the application of our ideas to those systems that are not known a priori to be critical. 

Contrary to plane and cylinder geometries, for the torus geometry we do not have any field theoretical results to guide our ansatz. 
We just know that close to the two entangling surfaces, the EH must be a linear function of the separation. 
A possible smooth interpolation between the two linear regimes is suggested by Eq. \eqref{BWtheoremCFT} which has a suitable generalization 
for a sphere in arbitrary dimension \cite{Casini:2011aa}. Following this line of thoughts, we propose the ansatz
\begin{eqnarray}
\Gamma_x &=& \frac{x\left( L/2 - x \right)}{L/2} \Gamma, \nonumber  \\
\Gamma_y &=& \frac{\left(x - \frac{1}{2}\right)\left[ \frac{L}2 - \left(x - \frac{1}{2}\right) \right]}{L/2} \Gamma, \nonumber \\
\Theta_{(x,y)} &=& \frac{\left(x - \frac{1}{2}\right)\left[ \frac{L}2 - \left(x - \frac{1}{2}\right) \right]}{L/2} \Theta.
\label{eq_dist1}
\end{eqnarray}

For the 1D critical case, exact EH profiles can be obtained by discretizing Eqs.\,\eqref{BWtheoremCFT2} and \eqref{BWtheoremCFT3}.
For the half-bipartition of length $L/2$ of the ring one has
\begin{eqnarray}
\Gamma_x &=& \frac{L}{2 \pi} \sin \( \frac{ 2 \pi x }{ L } \) \, \Gamma ,\nonumber  \\
\Theta_x &=& \frac{L}{2 \pi} \sin \(  \frac{ 2 \pi}{L} \( x - \frac{1}{2} \) \) \, \Theta ,
\label{eqdist2}
\end{eqnarray}
while for the open chain
\begin{eqnarray}
\Gamma_x &=& \frac{L}{\pi} \sin \(  \frac{ \pi x }{ L } \) \, \Gamma, \nonumber  \\
\Theta_x &=& \frac{L}{\pi} \sin \(  \frac{ \pi}{L} \( x - \frac{1}{2} \) \) \,  \Theta .
\label{eqdist3}
\end{eqnarray}

Finally, the overall energy scale in \eqref{BW-EH}, $\beta_{EH}$,
is related to the ``speed of light'', $v$,  in the corresponding low-energy field theory
\begin{equation}
 \beta_{EH} = \frac{2 \pi}{v}.
 \label{velocity}
\end{equation}
The reason to use the name $\beta_{EH}$ is that as for the thermodynamics ``beta'', $\beta = 1/T$ ($T$ is the temperature), the
BW overall energy scale plays the role of an effective temperature,
as will be discussed in next section.

The velocity $v$ may be fixed by matching the small momentum ($\hat p_k$) expansion of the lattice dispersion relation $E(k)$ with the
relativistic one $E(p)=\sqrt{m^2 v^4+ v^2 p^2}$. Such a velocity is generically different from the quasiparticle one $V(k)\equiv dE(k)/d p_k$. 
The two coincide only for gapless theories when $v=V(0)$, i.e.  the sound velocity.

\subsection{Regimes of applicability of the approach}

A natural question to ask is, to which extend field theory results on the functional form of the EH are applicable to lattice models 
and in which sense. The LBW EH is not generically an exact form, even in the thermodynamic limit. 
This is, e.g., explicitly manifest in free fermion results \cite{Eisler_2017,eisler2018} showing that the exact EH of a Fermi sea not only has tiny deviations compared to the field 
theoretical BW EH, but also presents very small longer range terms completely absent in \eqref{BWtheorem} 
(and the same happens also for the interacting XXZ spin chain~\cite{Nienhuis_2009}).
Conversely, for the gapped Ising chain, the LBW EH is exact \cite{peschel2009} even for very small correlation lengths - when lattice effects become dominant.  

Before discussing in the next subsection a series of quantitative criteria to determine the applicability regimes of the LBW EH 
  using numerical simulations (whose results are discussed in the next sections), we provide here a qualitative discussion. 

When transposing the field theory predictions above on finite lattice models, three ingredients shall be considered: 
{\it i)} the loss of Lorentz invariance due to the lattice, even when it is recovered as a low-energy symmetry; 
{\it ii)} for massive theories, the presence of a finite $\xi/a$ ratio, leading to potentially harmful UV effects at the lattice spacing level; 
{\it iii)} finite volume effects (which can be partially taken into account in 1D CFTs).

In quantum field theory language and close to a quantum phase transition, the loss of Lorentz invariance is typically attributed to the fact that the lattice turns on several irrelevant operators which directly affect the Hamiltonian spectrum. At the level of the EH, to the best of our knowledge, this has not been discussed so far. 
Since there is abundant evidence that universal properties of lattice models (such as the entanglement entropy of models described at low energies by CFTs~\cite{Calabrese_2009,laflorencie2016}) are in excellent agreement with field theory expectations, it is natural to argue that the microscopic EH is governed by the LBW EH, plus terms that depend on irrelevant operators. We note that, in the specific case of spin models, a set of recent ans\"atze proposed in Ref.~\onlinecite{alet2014} falls into this category. Under this assumption, it is possible to argue that low-lying entanglement properties should be well captured by the lattice BW EH at least in the critical case. Similar arguments are at the basis of the use of the ES in topological models~\cite{Swingle2012}, in particular for quantum Hall wave-functions. 

From a complementary viewpoint, it is possible to argue that, at least for the critical case, deviations are directly tied to curvature effects in the lattice dispersion relation. This sets an energy scale upon which excitations cease to be well described by a Lorentz invariant field theory. In the context of correlated fermions, we thus expect that the accuracy of the LBW EH degrades when the speed-of-light-to-band-width ratio becomes small - down to the flat band case, which is not expected to be captured at all. This expectation is confirmed by free fermions exact calculations \cite{Eisler_2017}.

The effects of a finite $\xi/a$ ratio have already been qualitatively discussed in Ref.~\onlinecite{Dalmonte:2017aa}: in brief, as long as the correlation length is not of the same order of the lattice spacing (thus making a field theory description not immediately applicable), these deviations are negligible. We note that, for what concerns the ES, it has been observed that in the massive regime of the Ising model~\cite{peschel2009}, in the close vicinity of the Affleck-Kennedy-Lieb-Tasaki point of bilinear-biquadratic spin-1 chains~\cite{Dalmonte:2017aa}, and in gapped XXZ spin-chains~\cite{Dalmonte:2017aa} the lattice BW EH is extremely accurate, so the validity of the approach even at $\xi\simeq a$ cannot be ruled out {\it a priori} (whilst has anyway to be justified {\it a posteriori}).

Finally, we discuss finite volume effects. Their estimate is non-trivial (except for those encoded in \eqref{eqdist2} and \eqref{eqdist3} for 1D CFTs): 
for this reason, we present below a finite-size scaling analysis of several quantities of interest. 
We anticipate that, at least for what concerns the low-lying entanglement spectrum, we observe universal scaling.

\subsection{Numerical checks of BW theorem on a lattice}

The reduced density matrix of subsystem $A$ is written in terms of the BW-EH as
\begin{align}
\rho_A \to \rho_{EH} = \frac{e^{-\tilde{H}_{A,BW}}}{Z},
\end{align}
where the constant $Z = \tr(e^{-\tilde{H}_{A,BW}})$, written in analogy to thermodynamics, ensures the normalization of $\rho_{EH}$.
For now on, we call the exact reduced density matrix,  $\rho_A$, and the one obtained with via the lattice BW, $\rho_{EH}$. The comparison of the thermal density matrix $\rho_{EH}$ and the exact one is addressed at  both the eigenvalues and eigenvectors level.

\paragraph{Entanglement spectrum.}
The first comparison between $\rho_A$ and $\rho_{EH}$ is at the level of the eigenvalues $\epsilon_\alpha$ of the corresponding entanglement Hamiltonian.
These eigenvalues are however affected  by the values of both the non-universal constant $c'$ in \eqref{BWtheorem} and of the entanglement temperature $\beta_{EH}$ in \eqref{BW-EH}.
These non-universal constants must be fixed either by an exact calculation or by an independent numerical study. 
In some cases in the following we will perform this direct analysis.
There is however an even better way to perform such a comparison  which does not require an a priori knowledge of these non-universal constants. 
Indeed, let us consider the ratios
\begin{equation}
\label{ratios}
 \kappa_{\alpha;\alpha_0} = \frac{\epsilon_{\alpha} - \epsilon_{0}}{\epsilon_{\alpha_0} - \epsilon_{0}},
\end{equation}
where $\epsilon_{0}$ is the lowest entanglement energy on the system (corresponding to the largest eigenvalue of $\rho_{A}$),
and $\epsilon_{\alpha_0}$ is a reference state suitably chosen to accomodate degeneracies of the lowest eigenvalue in the EH spectrum.
It is clear that the $c'$ dependence of the eigenvalues cancels out in the differences taken in the numerator and in the denominator in \eqref{ratios}. 
Taking the ratio in \eqref{ratios} cancels also the dependence on $\beta_{EH}$. 
For this reason we call the quantities \eqref{ratios} {\it universal ratios}.

 We use the Density Matrix Renormalization Group (DMRG) to obtain these quantities for quantum spin chains of length up to $100$ sites. The entanglement spectrum of the original system is computed keeping $100-150$ states and using the ground state as the target state in the proper symmetry sector. The lowest part of the BW-EH spectrum instead is obtained by targeting $5-10$ states in all the symmetry sectors. The magnitude of the discarded weight in the DMRG algorithm depends on the boundary conditions and on the system being homogeneous (exact ES computation) or not (BW-EH spectrum computation). When the homogeneous system has OBC/PBC we were able to keep the truncation error always below $10^{-12}$/$10^{-8}$ for the largest systems sizes considered. This is achieved in few DMRG sweeps, typically 2 or 3. All measurements were performed after a minimum of 5 sweeps to ensure convergence of the algorithm. Oppositely, in the inhomogeneous case, more sweeps were required for DMRG to converge, and a minimum of 6 sweeps was always performed before collecting the eigenvalues of the BW-EH. However, since the BW system is open, we were always able to keep the truncation error below $10^{-10}$ for all the chains considered in what follows~\footnote{This reflects the fact that entanglement eigenstates have always an entanglement content which is typically equal or smaller than that of the ground state wave function.}.  

\paragraph{Entanglement eigenvectors.}
In order to understand the accuracy of the BW EH at the eigenvector level, we consider the overlaps
\begin{equation}
\label{overlaps}
|\braket{\psi_{\alpha}^{EH}}{\psi_{\alpha'}^{A}}| = M_{\alpha,\alpha'} 
\end{equation}
for different levels of the spectrum. These eigenvectors are computed via Exact Diagonalization (ED) of both  $\rho_A$ and the BW-EH.

\paragraph{Correlation functions. }
Operators (observables) defined exclusively on subsystem $A$ are directly related to $\rho_{A}$ ($\rho_{EH}$)
\begin{equation}
\label{localoss}
 \left< O_A \right> = \tr(O_A \rho_{A})  \to \frac{\tr\big(e^{-\tilde{H}_A}O_A\big)}{Z}.
\end{equation}
Similarly the ground state properties of the subsystem $A$ are directly related to the \textit{thermal} properties of the EH-BW.
Hence, as another check of the BW construction we use the finite-temperature QMC method Stochastic Series Expansion (SSE) and finite-temperature DMRG\cite{white2005} to obtain local and two-body correlation functions of the BW-EH system. 
We then compare these quantities with the exact ground state expectation values computed via DMRG and QMC~\cite{sandvik1991,sandvik2002}.

The SSE method samples terms in a power series of $e^{-\tilde{H}}$ in the partition function using local and loop (directed loop) updates  \cite{sandvik2002}.
For the BW-EH system, as the local effective temperature decreases (Hamiltonian couplings increases) away from the boundary,
the use of loop updates is important to prevent the slowing down of autocorrelation times.
In fact, as shown in the supplemental material, the asymptotic autocorrelation times of local observables obtained with the directed-loop SSE algorithm  is much smaller then the typical number of QMC measurements that we use, $N_{meas} \approx 10^{8}$.
Thus, at least for the systems sizes that we consider ($L$ up to 100) the slowing down of autocorrelation times is not an issue for the SSE simulations of BW-EH.

Finite-temperature DMRG accuracy was checked by varying both the number of states kept during the imaginary time evolution and the Trotter step employed. Since the imaginary time evolution is applied on a state in which the system is maximally entangled with an ancilla, if the Hamiltonian conserves some quantum number one can exploit it by preparing the maximally entangled initial state within a given symmetry sector of the Hilbert space and restricting the evolution to that sector~\cite{white2005}. Using this technique we were able to reach convergence of the results by keeping a maximum of 150 states per block. We used first order Trotter decomposition, which means one Trotter step per half sweep, with a Trotter step of $10^{-3}$.

In the next two sections, we report our results on the three criteria above for a set of lattice models in one and two-dimensions. It is worth stressing how the three diagnostics employed are sensitive to different features of the reduced density matrix. Universal entanglement gap ratios are insensitive to possible errors in the prefactors of the entanglement Hamiltonian (i.e., to $\beta_{EH}$), and are not informative about eigenstates. Oppositely, overlaps between entanglement eigenvectors are not informative about the spectrum, but rather describe the accuracy in having the same eigenvectors. Finally, correlation functions are sensitive to all details of the EH - both spectra, correct speed of sound, and eigenvectors. However, they are also a somewhat less direct as a diagnostic - for instance, very close correspondence in correlation functions can be obtained by considering density matrices with very different eigenvectors.

\section{One-dimension}\label{sec_1d}
One-dimensional quantum systems represent an ideal framework to test the applicability of LBW EH predictions. The main advantage here is that wave-function based methods such as DMRG and ED can be pushed to considerably large system sizes. In addition, the CFT results of Ref.~\onlinecite{Cardy:2016aa} allow us to employ formulas which do consider a finite size of the subsystem (Eq.\,\eqref{BWtheoremCFT}) and of the system (Eq.\,\eqref{BWtheoremCFT2},\eqref{BWtheoremCFT3}), which implies that finite size effects can be controlled in a more efficient manner.

\subsection{Transverse Field Ising Model (TFIM)}
The quantum Hamiltonian of this model reads~\cite{fradkinbook}:
\be
H=-\sum_i \sigma^z_{i}\sigma^z_{i+1} -g\sum_i \sigma^x_i,
\ee
where $g>0$ and $\sigma^j$ are the Pauli matrices.
The model can be solved exactly and it is diagonalized in terms of spinless fermions (with mode operators $b^\dag_k,b_k$) as
\be 
H = \sum_k E(k) \( b^\dag_kb_k  -\frac{1}{2} \),
\ee
where $E(k) = \sqrt{ ( 1 - g )^2 + 4 g (\hat{p}_k)^2 } $, with $\hat{p}_k = \sin \pi k/L$ being the lattice momentum. 
By matching this dispersion relation with the relativistic one, we get the light velocity $v=2 \sqrt{g}$ as a function of the lattice parameter $g$.
The gap closes in the thermodynamic (TD) limit when $g=1$. A quantum phase transition occurs at this point, separating a ferromagnetic phase for $g<1$ from a paramagnetic phase for $g>1$. In the former the $\Z_2$ symmetry of the model is broken by the ground state of the system, which is degenerate in the TD limit.
The low energy physics of the quantum critical point is described by a $c=1/2$ CFT.

For this model we expect the lattice discretization of Eq. \eqref{BWtheorem} (i.e. Eq.~\eqref{eq_dist}) to work well for a chain with OBC as long as the correlation length in the system is large w.r.t. to the lattice spacing and small compared to the system size. In fact, the EH for a half-partition 
of an infinite chain can be computed exactly in the coordinate basis away from the critical point~\cite{peschel2009}. The result perfectly matches our lattice version of BW-theorem, although it does not predict the prefactor $\beta_{EH}$. 
In the PBC case instead we expect conformal BW-theorem Eq.\,\eqref{BWtheoremCFT2} to fail as soon as a gap opens in the energy spectrum. 

\begin{figure}[]
{\centering\resizebox*{8.7cm}{!}{\includegraphics*{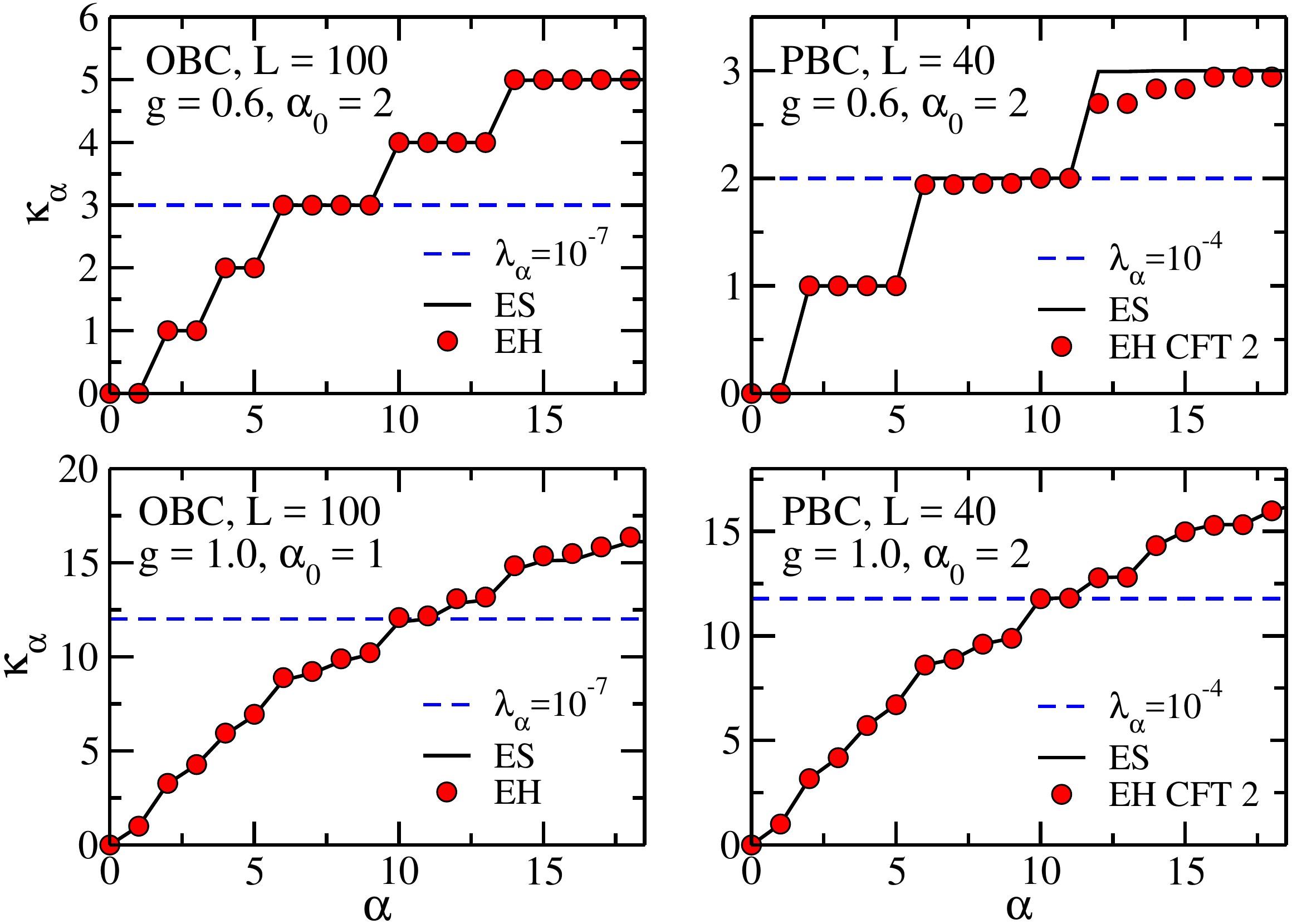}}}
\caption{Ratio $\kappa_\alpha$s for the transverse field Ising chain. The black solid line and red circles stand for ratios computed from the exact ES and the EH spectrum respectively. The blue dashed line marks $\rho_A$ eigenvalues with a magnitude indicated in the legend. 
PBC data slightly deviate from the field theory prediction in the ferromagnetic gapped phase $g=0.6$, the maximum relative error being larger that 3\% after the 13th eigenvalue.} 
\label{is_fig_spec} 
\end{figure}

Fig.\,\ref{is_fig_spec} shows the universal ratios Eq.\,\eqref{ratios} computed from the ES both assuming OBC and PBC (black solid line). 
These ratios are compared to the ones computed from the LBW-EH Eq. \eqref{eq_dist} in the former case and to the ones computed from Eq. \eqref{eqdist2} in the latter case (red circles). At the critical point the agreement is almost perfect in both cases: relative errors of the ratios are always smaller than 2\%. Instead, in the ferromagnetic gapped phase, slight discrepancies are observed when the system is subjected to PBC: the ratios agree within 3\% only as long as $\lambda_\alpha\lesssim 10^{-4}$.

\begin{figure}[]
{\centering\resizebox*{8.7cm}{!}{\includegraphics*{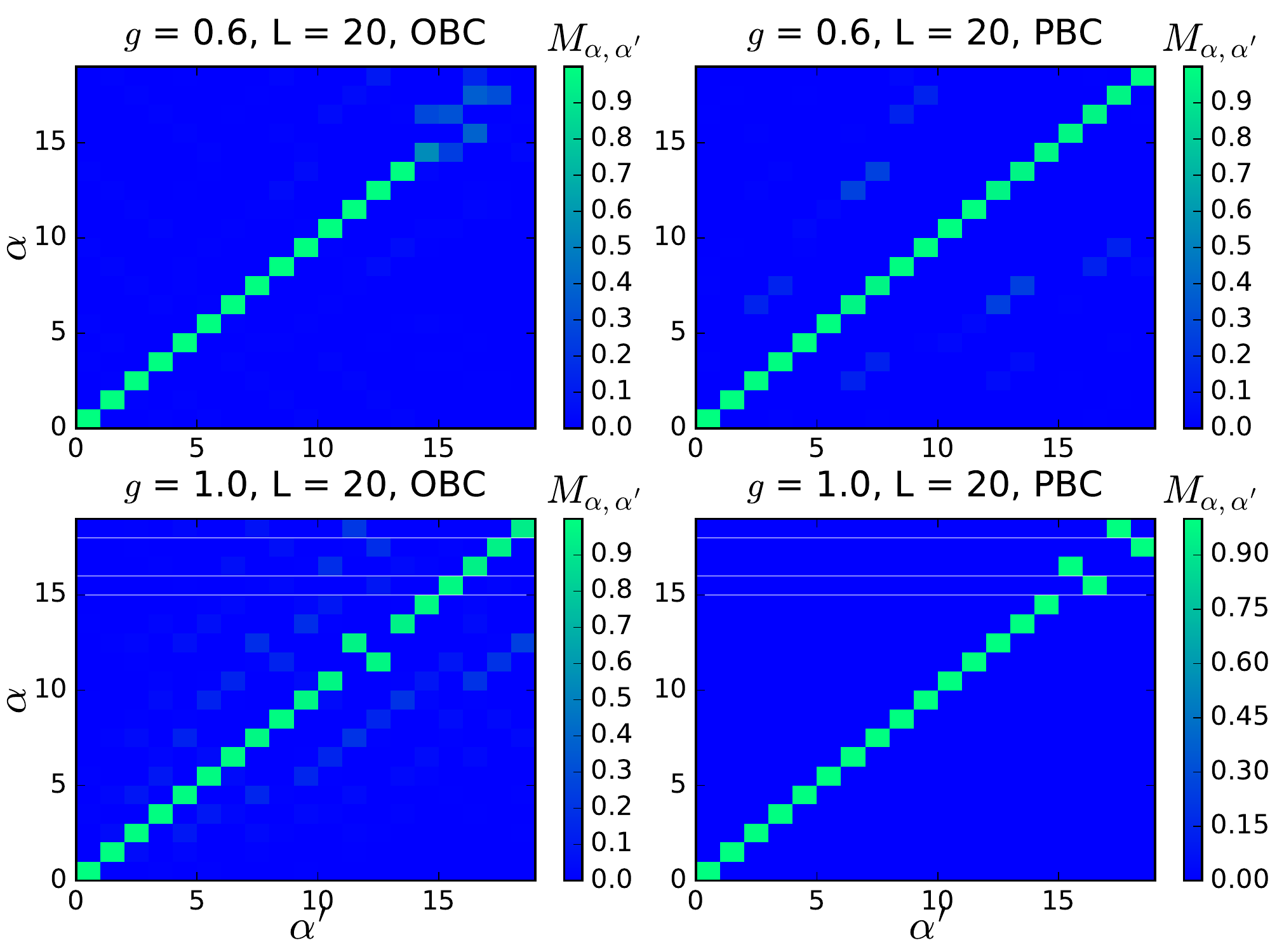}}}
\caption{Overlaps as defined in Eq.\,\eqref{overlaps} for the transverse field Ising chain in the ferromagnetic phase $g=0.6$ and at the critical point $g=1.0$.
Deviations from unity on the diagonal are of order $10^{-3}$ in all the cases considered up to the first 13 eigenstates. The few points close to the diagonal correspond to exact degeneracies in spectrum (not all degeneracies are off-set).} 
\label{is_fig_ov} 
\end{figure}

Moving to the eigenvectors, 
the overlaps in Eq.\,\eqref{overlaps}, computed via ED, are plotted in Fig.\,\ref{is_fig_ov}. Both in the OBC and PBC cases the magnitude of the overlaps is 1 with $10^{-3}$ accuracy, independently of the system being critical or gapped. Note however that overlaps of order $10^{-1}$ are observed also away from the diagonal at the critical point in the OBC case and when the system is gapped in the PBC case. The latter fact is expected since Eq.\,\eqref{BWtheoremCFT2} should provide the EH of a gapless system.
We did also check the finite size scaling of the matrix norm of the difference between $\rho_A$ and $\rho_{EH}$ i.e. $|| \rho_A - \rho_{EH} ||$, where $||A|| = \sqrt{ \tr( A A^\dagger ) }$. The magnitude of the matrix norm is of order $10^{-2}$ for the system sizes accessible with ED and it decreases with system size in all the cases considered.

\begin{figure}[b]
{\centering\resizebox*{8.7cm}{!}{\includegraphics*{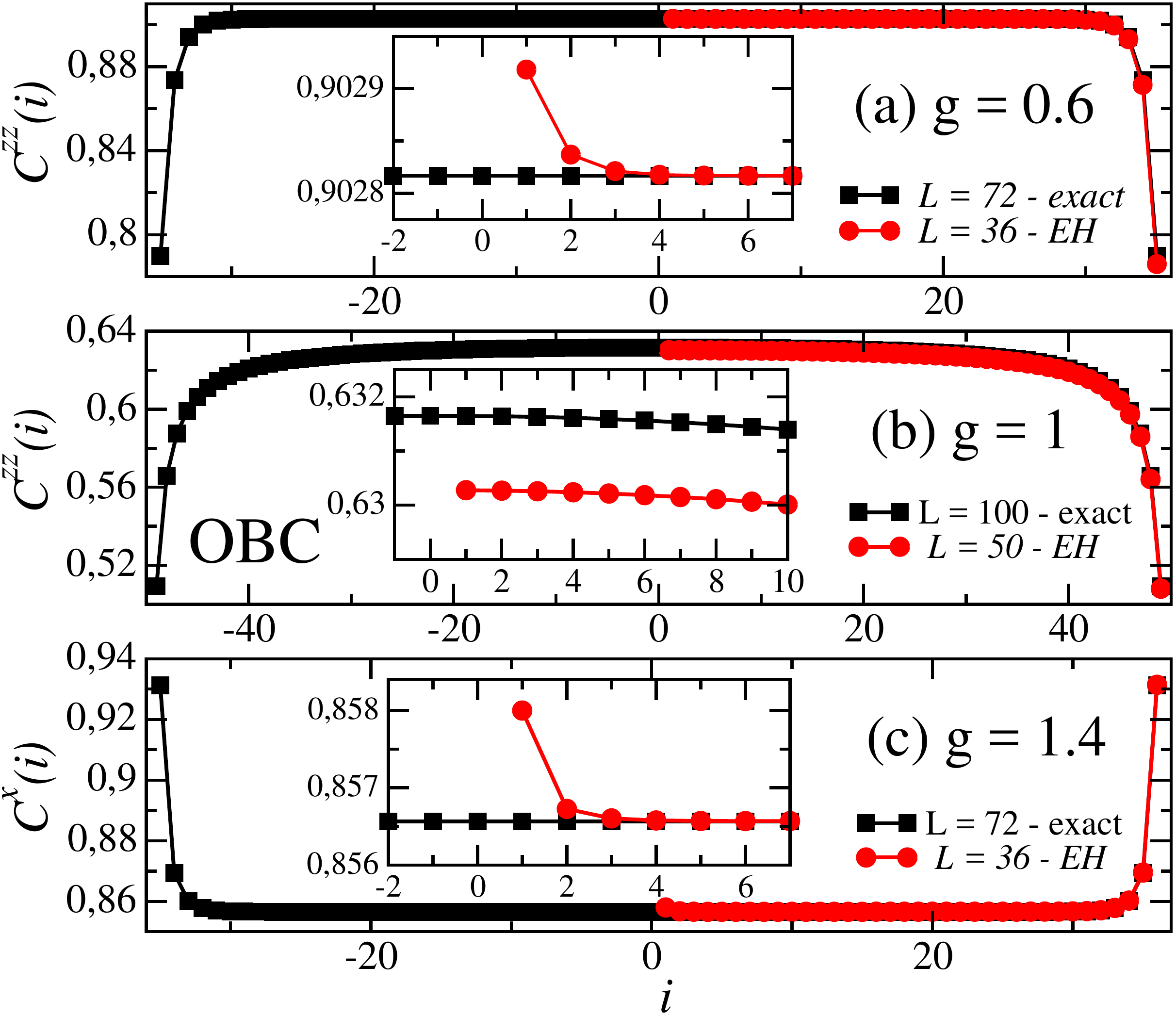}}}
\caption{(a-b): Local correlation function as defined in Eq.\,\eqref{ising_corr} for the transverse field Ising chain in the ferromagnetic phase $g=0.6$ and at the critical point $g=1.0$.
The square (black) and circle (red) points are results for the  original and the half-bipartion EH-BW systems, respectively. (c): local transverse magnetization in the paramagnetic phase.} 
\label{is_fig_corr} 
\end{figure}

\begin{figure}[]
{\centering\resizebox*{8.3cm}{!}{\includegraphics*{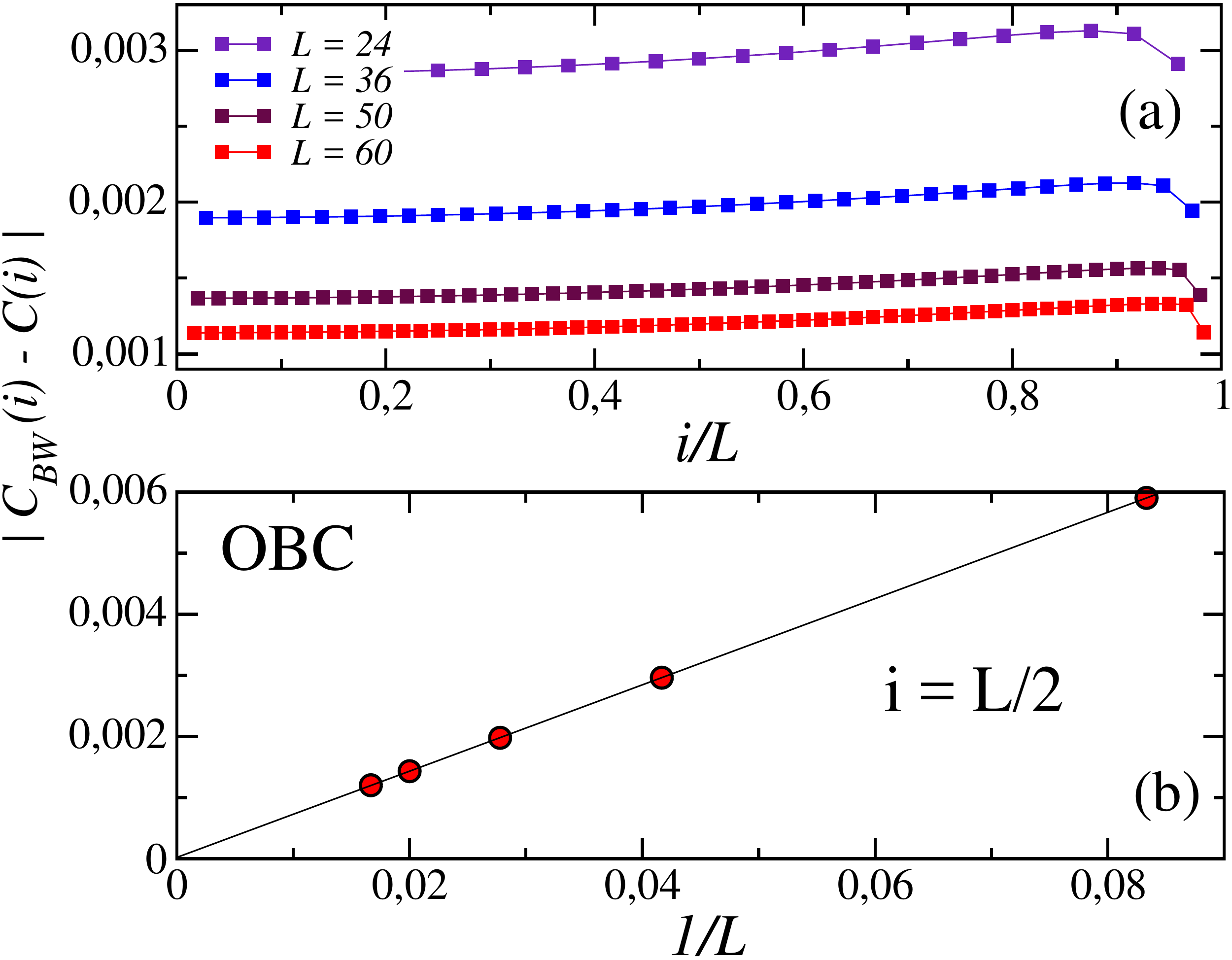}}}
\caption{Finite size scaling of the difference between BW thermal- and ground state- expectation values at the critical point of the Ising model, with OBCs. In (a), deviations are plotted for all the sites in the subsystem and they are largest close to the boundary away from the cut. In (b), deviations are plotted for a site in the middle of the subsystem and they clearly scale to zero as a power-law. } 
\label{is_corr_scal} 
\end{figure}

Expectation values of local observables are the only quantities considered here which are sensitive to the entanglement temperature. They thus probe more in depth this specific aspect of the BW theorem, which states that $\beta_{EH} = 2 \pi/v$. Thanks to the exact solution of the TFIM we know that $v = 2 \sqrt{g}$. 
The local observable we consider for this model is
\begin{equation}
 C^{zz}(i) = \left< \sigma_{i}^z \sigma_{i+1}^z\right>.
 \label{ising_corr}
\end{equation}
Note that, since the two points are nearest-neighbours, this observable is expected to be the most sensitive to finite-lattice-spacing effects.
The result of the comparison is depicted in Fig.\,\ref{is_fig_corr}(a-b) for the OBC case. 
The EH-BW results (red circles) are obtained as thermal averages  computed via finite-temperature DMRG. Ground state averages (black square) are obtained using DMRG with the ground state of the system as a target state. The agreement is excellent (below percent level) in the gapped paramagnetic phase even close to the cut, where the choice of the proper $\beta_{EH}$ almost completely cancels boundary effects. 
Relative errors in the bulk (including the open (right) boundary) are uniformly of order $10^{-6}$, while they reach a magnitude of $10^{-3}$ close to the cut (see inset). 
At the critical point instead we observe uniform deviations of 0.5\% over the whole half-chain. These are caused by finite size effects.
Fig.\,\ref{is_corr_scal} shows the difference between the thermal LBW expectation value and the ground-state one for different system sizes. 
Discrepancies exhibit power-law scaling to zero. 

In addition, we have also considered the expectation value of the transverse magnetization (i.e. along the $x$-axis)
\begin{equation}
 C^x(i) = \left< \sigma_{i}^x\right>.
 \label{ising_corr2}
\end{equation}
In Fig.~\ref{is_fig_corr}c, we show the corresponding spatial profile under OBCs: the behavior is very similar to that of the $C^{zz}$ correlator, with the maximum deviations of order $10^{-4}$ close to the boundary.

\subsection{Quantum three-state Potts Model (3PM)} \label{3pm}

The quantum Hamiltonian of the three-state Potts Model is given by~\cite{MussardoBook}:
\be 
H=-\sum_i \left(\sigma_i\sigma_{i+1}^{\dagger}+\sigma_{i}^{\dagger}\sigma_{i+1}\right)-g\sum_i \left(\tau_{i}+\tau_{i}^{\dagger} \),
\label{H3pm}
\ee
where $g>0$. The $\sigma$ and $\tau$ matrices are defined as
\be 
\sigma \ket{ \gamma } = \omega^{\gamma - 1 } \ket{\gamma},
\quad \quad 
\tau \ket{ \gamma } = \ket{ \gamma + 1 },
\vspace{2mm}
\quad \quad \omega = e^{i 2 \pi/3 },
\ee
and 
$\gamma = 0,1,2$. 

The phase diagram of this quantum chain is analogous to the TFIM one. 
The symmetry of the model is $\mathbb{Z}_3$ which is broken in the ferromagnetic phase with three degenerate ground states.  
Another important difference w.r.t. to the TFIM is that the Hamiltonian Eq.~\eqref{H3pm} is non-integrable away from the critical point at $g=1$. 
Here the spectrum can be computed \cite{kedem1994} in terms of massless excitations 
whose dispersion relation reads
\be \label{potts_disp}
E(k) = 
\frac{ 3 \sqrt{3} }{2 } \, \hat{p}_k\,, 
\ee
which matches the massless relativistic one with a sound velocity $v= 3 \sqrt{3}/2$. This critical point  is  described by a CFT with central charge $c=4/5$.~\cite{CFT1997}

Lorentz invariance is expected for the continuum limit of the lattice theory also away from the gapless conformally invariant point and thus the discrete BW theorem Eq.\,\eqref{eq_dist} is expected to hold also when $g \ne 1$.

\begin{figure}[]
{\centering\resizebox*{8.7cm}{!}{\includegraphics*{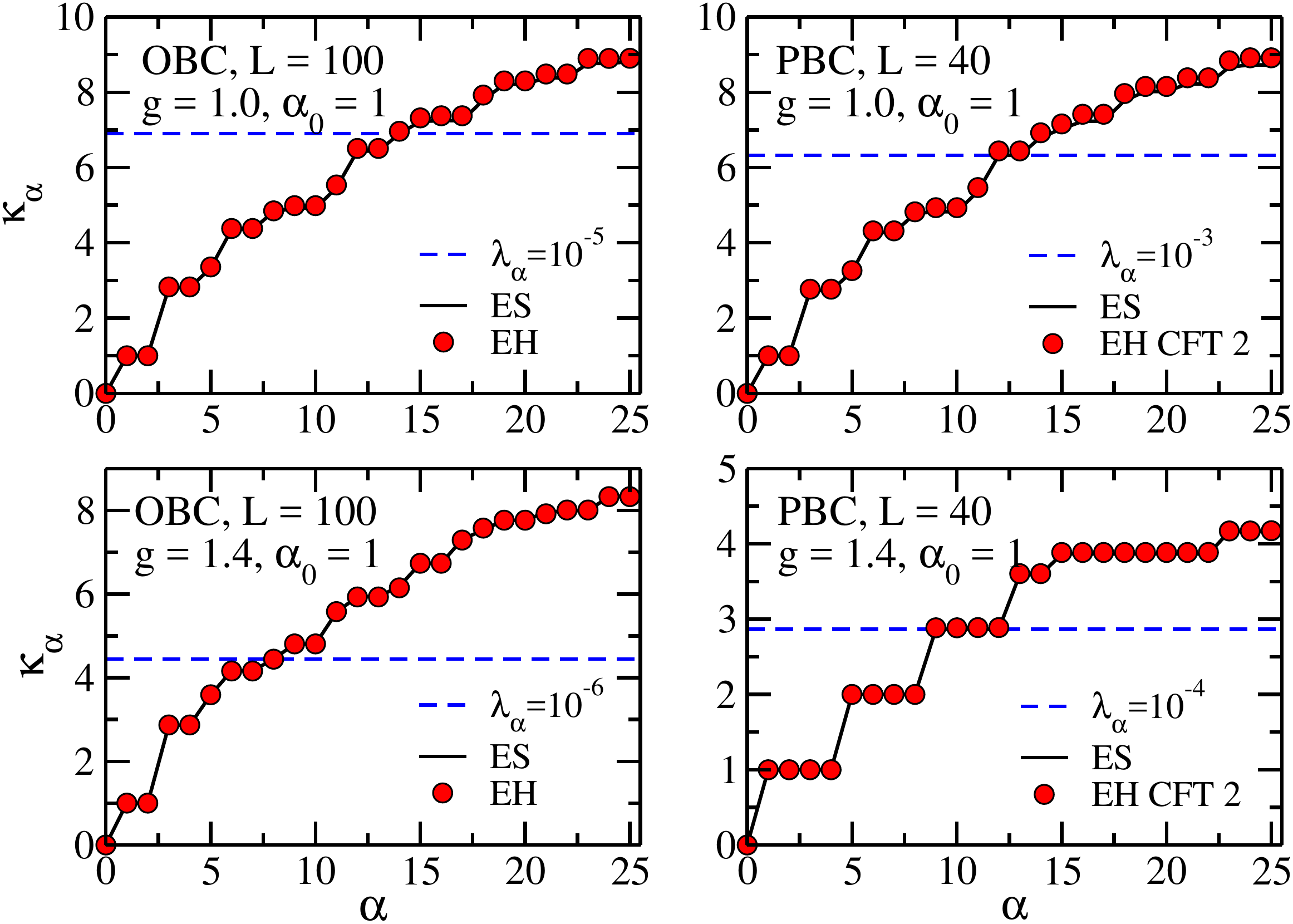}}}
\caption{Ratio $\kappa_\alpha$s for the 3-state Potts chain. The black solid line and red circles stand for ratios computed from the exact ES and the EH spectrum respectively. The blue dashed line marks $\rho_A$ eigenvalues with a magnitude indicated in the legend.} 
\label{po_fig_spec} 
\end{figure}

Fig.\,\ref{po_fig_spec} shows the comparison between the universal ratios Eq.\,\eqref{ratios} obtained from the ES and from the BW-EH for the system at the critical point ($g=1$) and in the paramagnetic phase ($g=1.4$). We see good agreement for both OBC and PBC also away from the critical point. 
In particular relative errors for the first 26 eigenvalues are smaller than 2\% at the critical point, in both the OBC and PBC case. In the gapped paramagnetic phase instead their maximum magnitude is 0.5\% and 4\% in the OBC and PBC case respectively.

\begin{figure}[]
{\centering\resizebox*{8.7cm}{!}{\includegraphics*{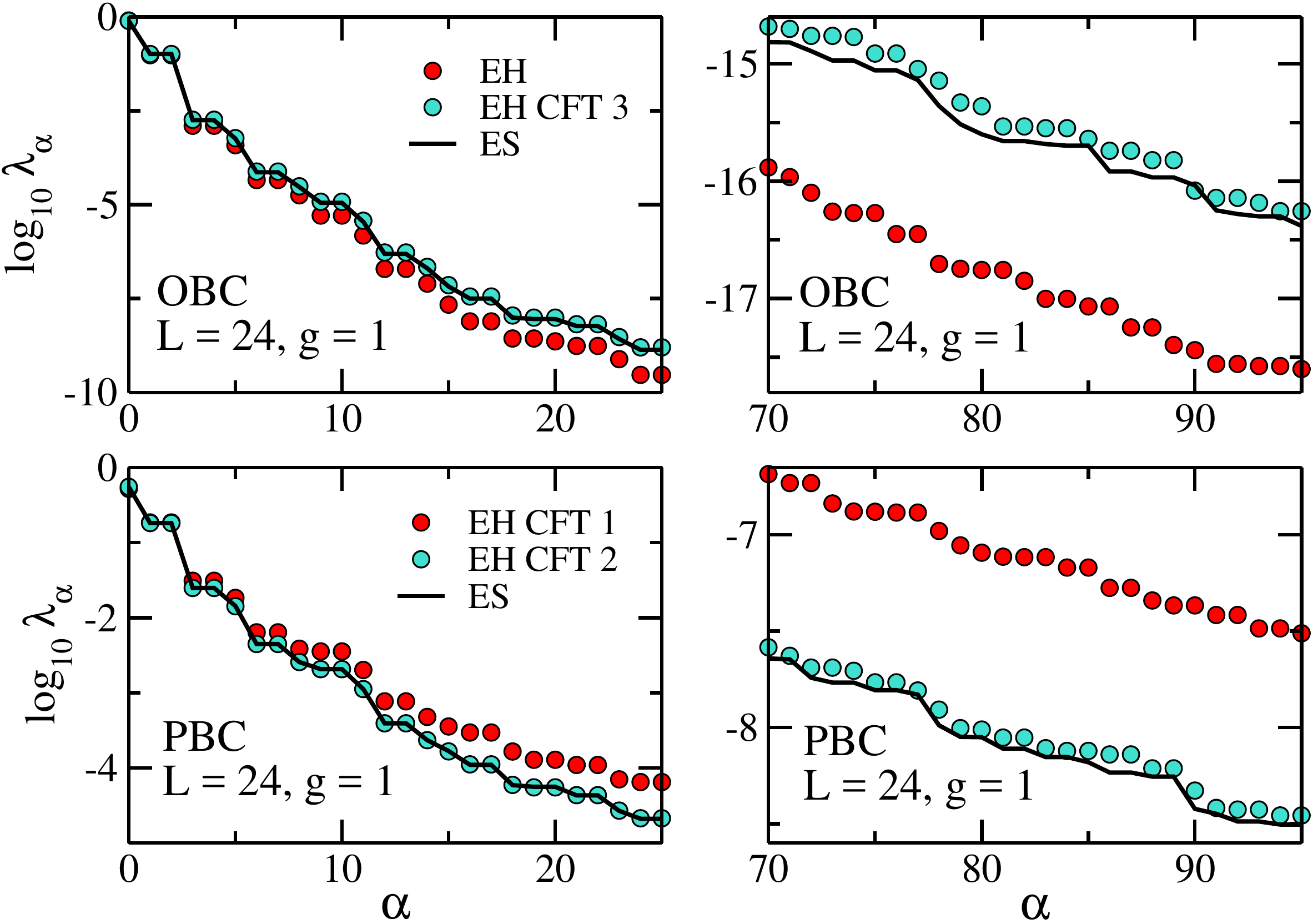}}}
\caption{Spectra comparison for the 3-state Potts chain at the critical point. The black solid line is the exact ES. Red and turquoise circles are the $\rho_{EH}$ eigenvalues computed via the infinite system EH (Eq.\eqref{BWtheorem} for OBC and Eq.\,\eqref{BWtheoremCFT} for PBC) and via the CFT finite system EH (Eq.\eqref{BWtheoremCFT3} for OBC and Eq.\,\eqref{BWtheoremCFT2} for PBC) respectively. On the left $\lambda_\alpha$s with $\alpha=0,\dots 25$ and on the right with $\alpha=70,\dots,95$.} 
\label{po_fig_spec2} 
\end{figure}

For this model we performed also a direct comparison of the spectra of $\rho_A$ and $\rho_{EH}$ at the critical point for which we need the sound velocity in Eq.\,\eqref{potts_disp}. 
For OBC, Fig. \ref{po_fig_spec2} shows the ES obtained by using both the infinite system EH \eqref{eq_dist} and the finite size CFT \eqref{eqdist3}.
The discrepancies between the ES of a finite system and the ES obtained from the infinite system EHs completely disappear when the CFT finite system EHs are used.
For PBC instead the lattice discretization of the conformal EH BW in Eq. \eqref{eqdist2} is used which matches the direct results perfectly. For the sake of comparison, we have also computed the ES using with Eq.~\eqref{eq_dist1}: the results, while approximately matching the density of states of the original model, are not able to reproduce the ES quantitatively. This comparison boosts the predictive power of the correct CFT EH, which, even on the lattice, almost completely suppressed finite size effects.

\begin{figure}[]
{\centering\resizebox*{8.7cm}{!}{\includegraphics*{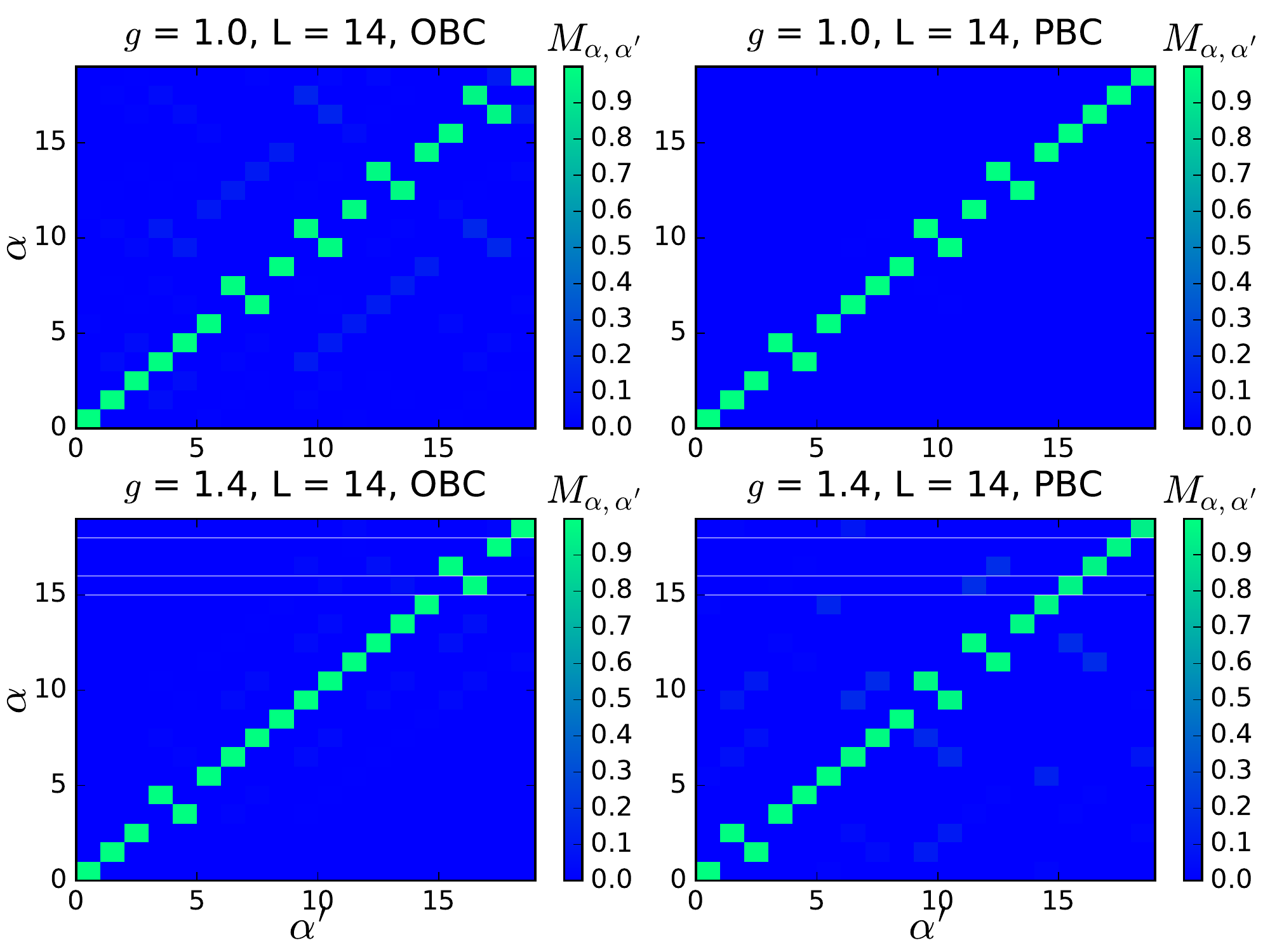}}}
\caption{Overlaps as defined in Eq.\,\eqref{overlaps} for the three-state Potts chain at the critical point $g=1.0$ and in the paramagnetic phase $g=1.4$. Deviations from unity on the diagonal are smaller than $10^{-3}$. The few points close to the diagonal correspond to exact degeneracies in spectrum. 
Largest non-vanishing overlaps away from the diagonal are observed in the PBC case when the gap spoils conformal invariance and thus the validity of Eq.\,\eqref{BWtheoremCFT2}.} 
\label{po_fig_ov}
\end{figure}

\begin{figure}[]
{\centering\resizebox*{8.7cm}{!}{\includegraphics*{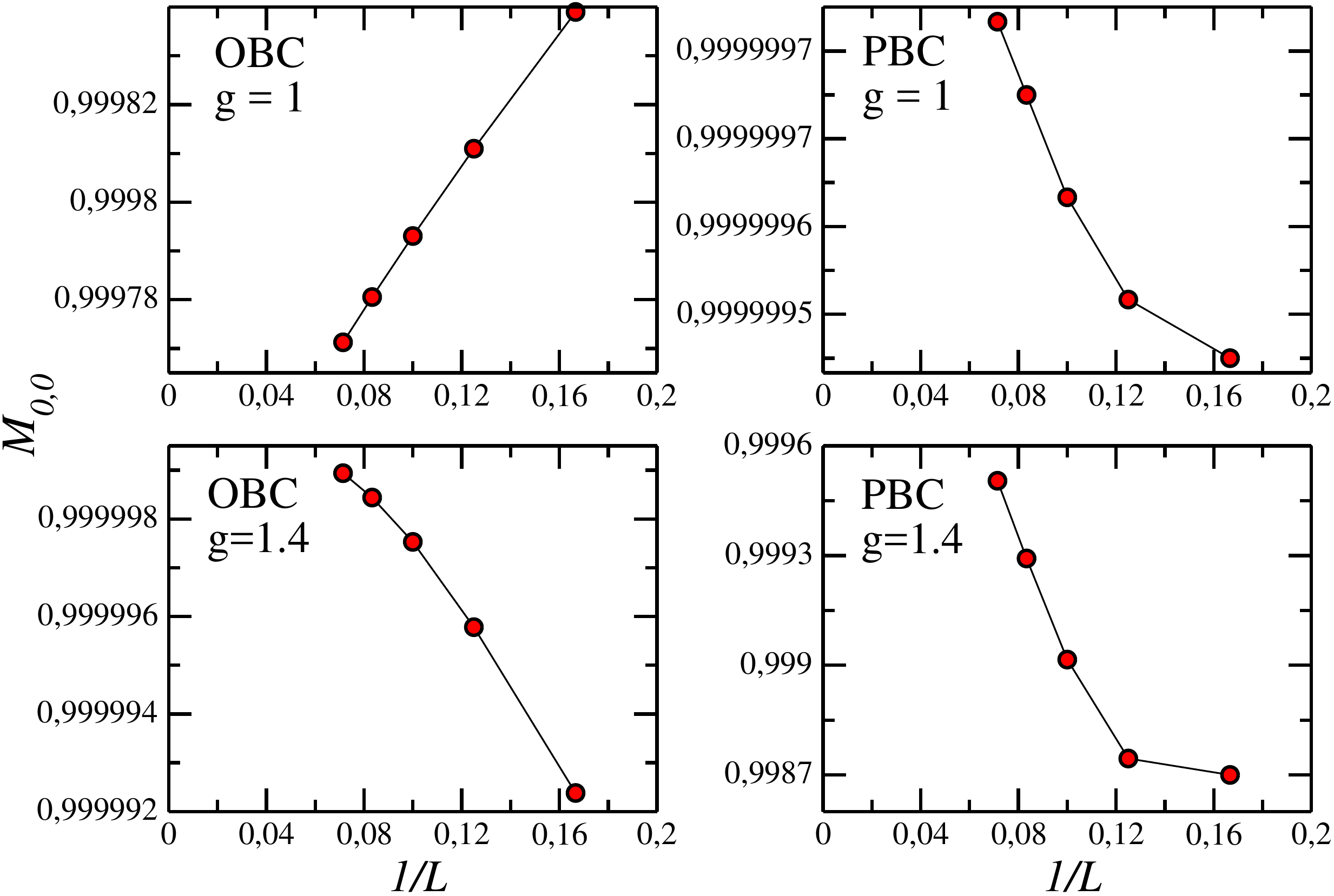}}}
\caption{Finite size scaling of the ground state overlaps $M_{0,0}$ as defined in Eq.\,\eqref{overlaps} for the three-state Potts chain at the critical point $g=1.0$ and in the paramagnetic phase $g=1.4$. } 
\label{po_scal_ov}
\end{figure}

Overlaps, computed via ED, between $\rho_A$ and LBW-EH eigenvectors are shown in Fig.\,\ref{po_fig_ov}, where we can observe the very same outcome as in the TFIM. Large ($\geq 1-10^{-3}$) overlaps involve all the first states in the two spectra both in the OBC and PBC cases. 
Non vanishing overlaps away from the diagonal are observed only in the gapped phase when the CFT EH Eq.\,\eqref{BWtheoremCFT2} is employed, as expected. In Fig.\,\ref{po_scal_ov} we report also the finite size scaling of the lowest overlap $M_{0,0}$, which decreases/increases when the system is gapless/gapped. The apparent decreasing behaviour of the overlap at the critical point might be an artifact of the small system sizes accessible with ED for this model. A trustworthy extrapolation to the thermodynamic (TD) limit is not possible: however, it is very indicative that changes over a large window of $L$  are at most of order $10^{-4}$, strongly suggesting that the overlap will remain finite in the TD limit - a remarkable fact given that we are looking at eigenvector properties. We have obtained similar results for all the 1D models discussed in this paper, but we did not report them because they are qualitatively equivalent to those in  Fig.~\ref{po_scal_ov}.

We finally consider the two-point function of the order parameter at neighbour sites:
\begin{equation}
 C(i) = \left< 2\, \mathrm{Re} \, \sigma_{i} \sigma^\dagger_{i+1}\right>.
 \label{po_corr}
\end{equation}
We compute this correlation function only for $g=1$ in the OBC case because the sound velocity is known exactly only at the critical point. In order to use an unbiased approach here, which does not rely on the CFT know-how of finite volume effects embodied in Eq.~\eqref{BWtheoremCFT3}, we have utilized the original BW formulation.

\begin{figure}[]
{\centering\resizebox*{8.7cm}{!}{\includegraphics*{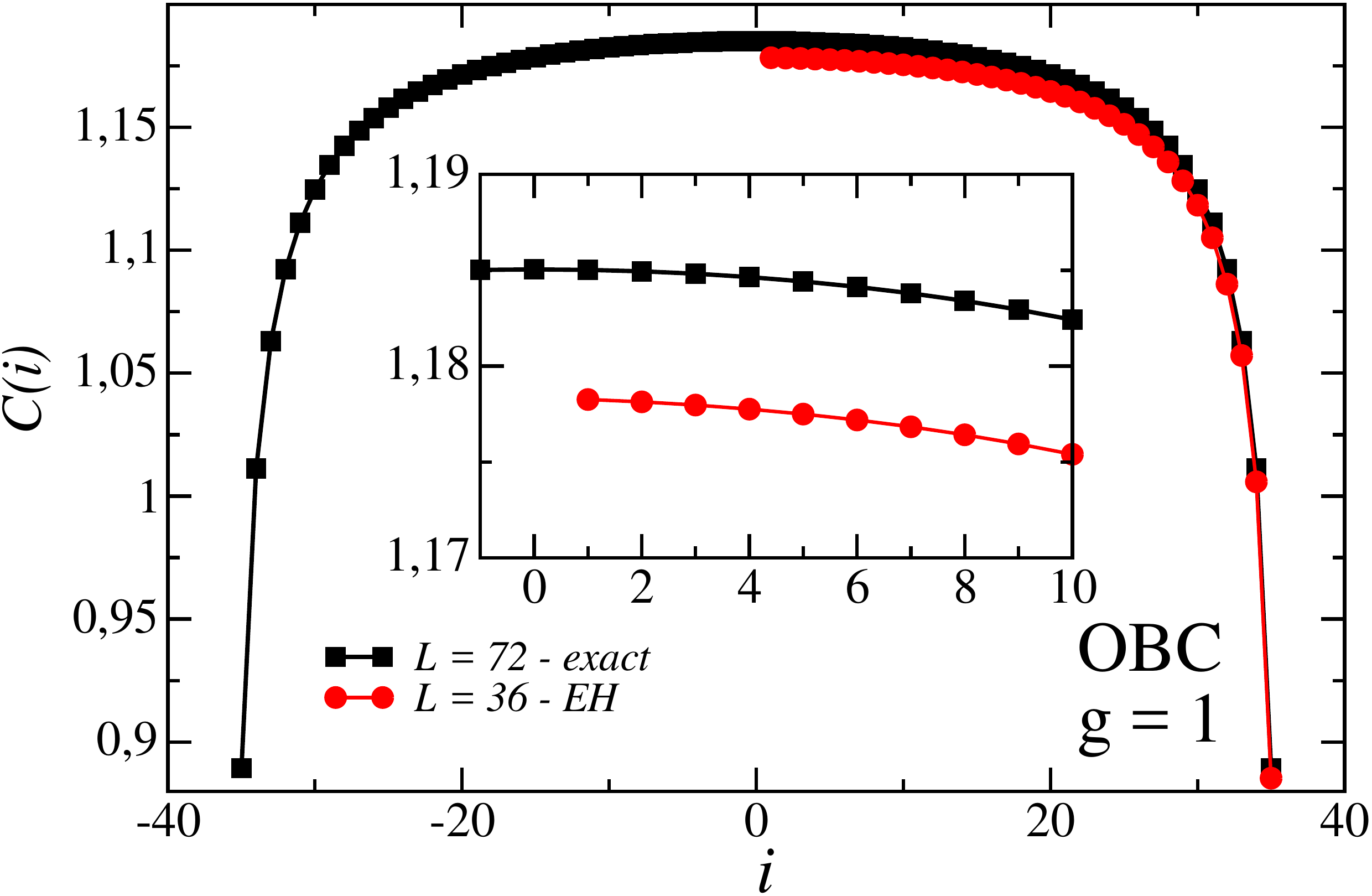}}}
\caption{Nearest-neighbor correlation function  Eq.\,\eqref{po_corr} for the three-state Potts chain at the critical point $g=1$.
The square (black) and circle (red) points are results for the exact and the half-bipartition LBW EH systems, respectively. The inset magnifies the region close to the boundary.} 
\label{po_fig_corr} 
\end{figure}

We used finite-temperature DMRG for the BW thermal average and ground state DMRG for the pure average over the ground state of the system. The result is reported in Fig.\,\ref{po_fig_corr}. As in the Ising case $0.5\%$ discrepancies are observed uniformly on the whole subsystem length. As they reduce considerably when increasing system size, we attribute their origin to finite volume effects as in the TFIM case.

\subsection{$S=1/2$ XXZ Model}\label{xxzsec}
The Hamiltonian of the XXZ spin-$1/2$ chain  is~\cite{fradkinbook}:
\be \label{xxz}
H=\sum_i \( S^x_i S^x_{i+1} + S^y_i S^y_{i+1} + \Delta S^z_{i} S^z_{i+1}  \).
\ee
This model is exactly soluble via Bethe ansatz, and its phase diagram supports three distinct phases. It is ferromagnetic for $\Delta <-1$, gapless critical (Luttinger liquid) for $-1 <\Delta \le 1$ and antiferromagnetic for $\Delta > 1$. In the ferromagnetic phase, the $\Z_2$ spin reversal symmetry is spontaneously broken. The critical phase is described by a $c=1$ CFT with varying Luttinger parameter $K =  \pi / 2 \arccos(-\Delta)$. The antiferromagnetic phase exhibits non-zero staggered magnetization, thus the spin reversal symmetry is broken by the 2-degenerate (quantum dressed) N\'eel states which live in the $S_z^{\rm tot} = 0$ sector.

For $\Delta < -1$ the low-lying excitations above the two magnetized ground states are translational invariant combinations of single-spin-flip states (magnons). Their exact dispersion relation reads
\be
E(p_k ) = 2 \( 1 - \cos \Big(\frac{2 \pi k }{L }\Big) - ( \Delta + 1 ) \),
\ee
which does not become relativistic in the continuum limit. At $\Delta=-1$ the gap closes but the magnon dispersion  remains quadratic. 
Thus there is no underlying Lorentz invariance  for $\Delta \le -1$. 

In the critical phase instead CFT predictions are  in perfect agreement with lattice results for what concerns spectral~\cite{Alcaraz_1987} as well as correlation function properties~\cite{MussardoBook,korepinbook,Kitanine_2014}. Therefore, we expect the LBW theorem to be accurate in this phase. The point $\Delta = 1$ hosts a BKT phase transition which links the AFM phase to the critical line. Close to this point the fundamental excitations are usually called spinons and their dispersion relation is
\be \label{spinon}
E(k) = \frac{\pi}{2} \sin \(  \frac{2 \pi k}{L} \) = \frac{\pi}{2} \, \hat{p}_k\,, 
\ee
from which we read the sound velocity $v = \pi/2$ by comparison with the massless relativistic dispersion relation. 
Indeed, the sound velocity is exactly known in the entire  critical line~\cite{korepinbook}:
\be \label{vxxz}
v = \frac{ \pi \sqrt{1-\Delta^2} }{ 2 \arccos \Delta }.
\ee
In the N\'eel phase the quasi-particles acquire a mass, but in the scaling region close to $\Delta \to 1^+$ they do not to spoil relativistic invariance of the continuum theory. We point out that some of the results discussed here for OBC are connected with Refs.~\onlinecite{Alba:2012aa,Kim_2016,mbeng-17}, which investigate the comparison between ES and the corner transfer matrix.

\begin{figure}[]
{\centering\resizebox*{8.7cm}{!}{\includegraphics*{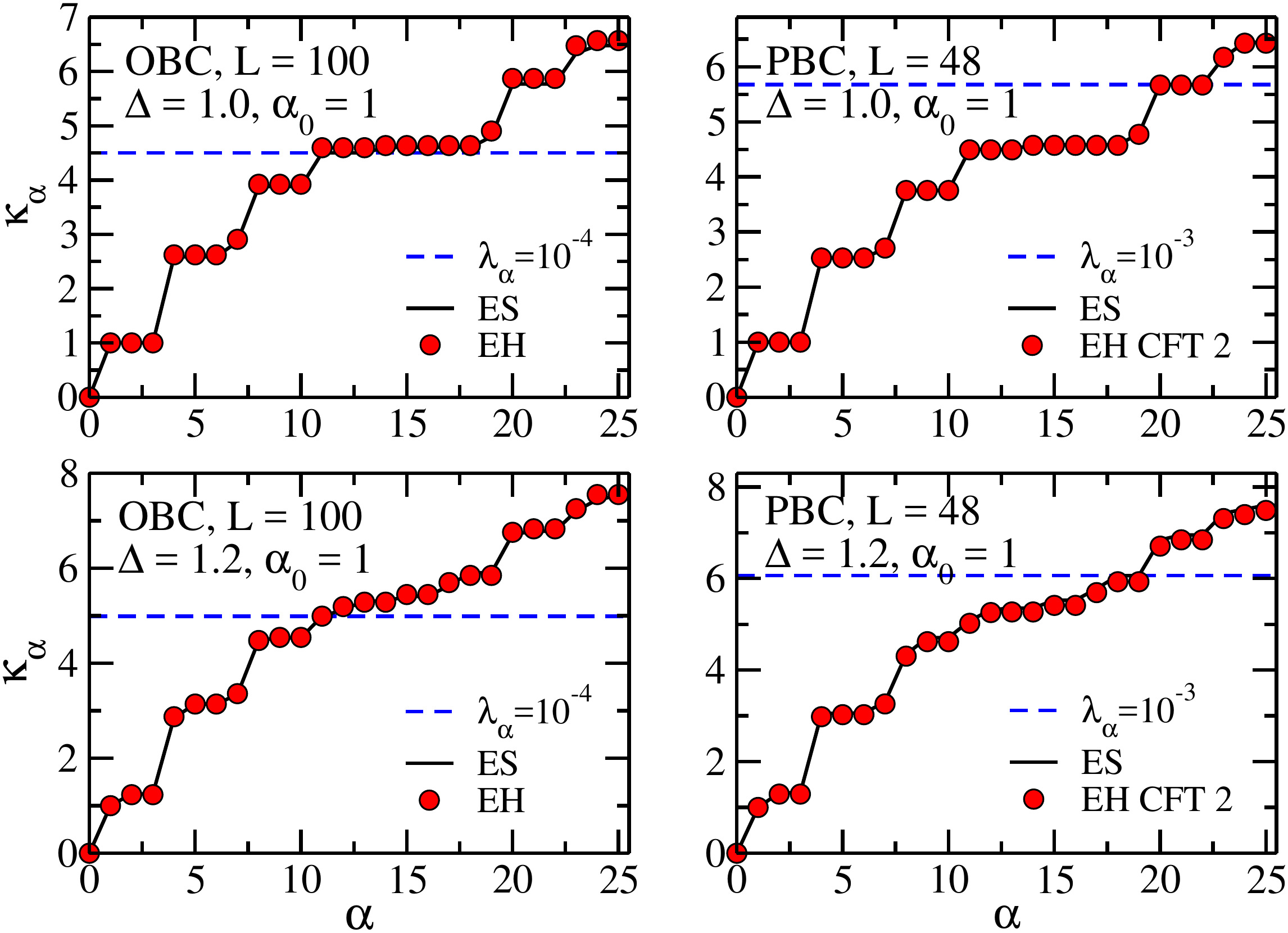}}}
\caption{Ratio $\kappa_\alpha$s for the XXZ spin $1/2$ chain. The black solid line and red circles stands for ratios computed from the exact ES and the EH spectrum respectively. The blue dashed line marks $\rho_A$ eigenvalues with a magnitude indicated in the legend. 
PBC data slightly deviate from the field theory prediction in the anti-ferromagnetic gapped phase $\Delta=1.2$, where 5\% discrepancies are observed among the lowest 20 eigenvalues. } 
\label{xxz_fig_spec}
\end{figure}

The comparison for the universal ratios \eqref{ratios}  between $\rho_A$ and BW-EH spectra is reported in Fig.\,\ref{xxz_fig_spec} for BKT point $\Delta=1$ and close to it in the gapped AFM phase $\Delta=1.2$. We note that analysis of these ratios for different parameter regime and the same model was presented in Ref.~\onlinecite{Dalmonte:2017aa}. We choose as a representative of the critical phase the transition point, where (logarithmic) finite size corrections are known to be the largest~\cite{Eggert:1996aa,Dalmonte:2011aa}.
As discussed in the previous sections the largest deviations are observed in the PBC case when the system is massive, where they reach 5\% also within the first 20 eigenvalues. In the massive OBC case instead they are smaller than 4\% for all the 30 eigenvalues considered. Maximum relative errors for $\Delta = 1$ are 2\% for both OBC. We have verified that in the lowest part of the spectrum, the difference of the universal ratios scales to 0 in the TD limit as a power law (the corresponding exponent needs further studies to be accurately determined).

As in Sec.\,\ref{3pm}, we report a direct comparison of the spectra of $\rho_A$ and $\rho_{EH}$ at the BKT point $\Delta=1$, exploiting the knowledge of the sound velocity \eqref{vxxz}. 
Again for OBC we use both the infinite size formula \eqref{eq_dist} and the CFT finite size one \eqref{eqdist3}.
The latter perfectly reproduces the exact data.
For PBC we only employ the lattice discretization of the CFT formula \eqref{eqdist2} finding a perfect match with the data from $\rho_A$. 

\begin{figure}[]
{\centering\resizebox*{8.7cm}{!}{\includegraphics*{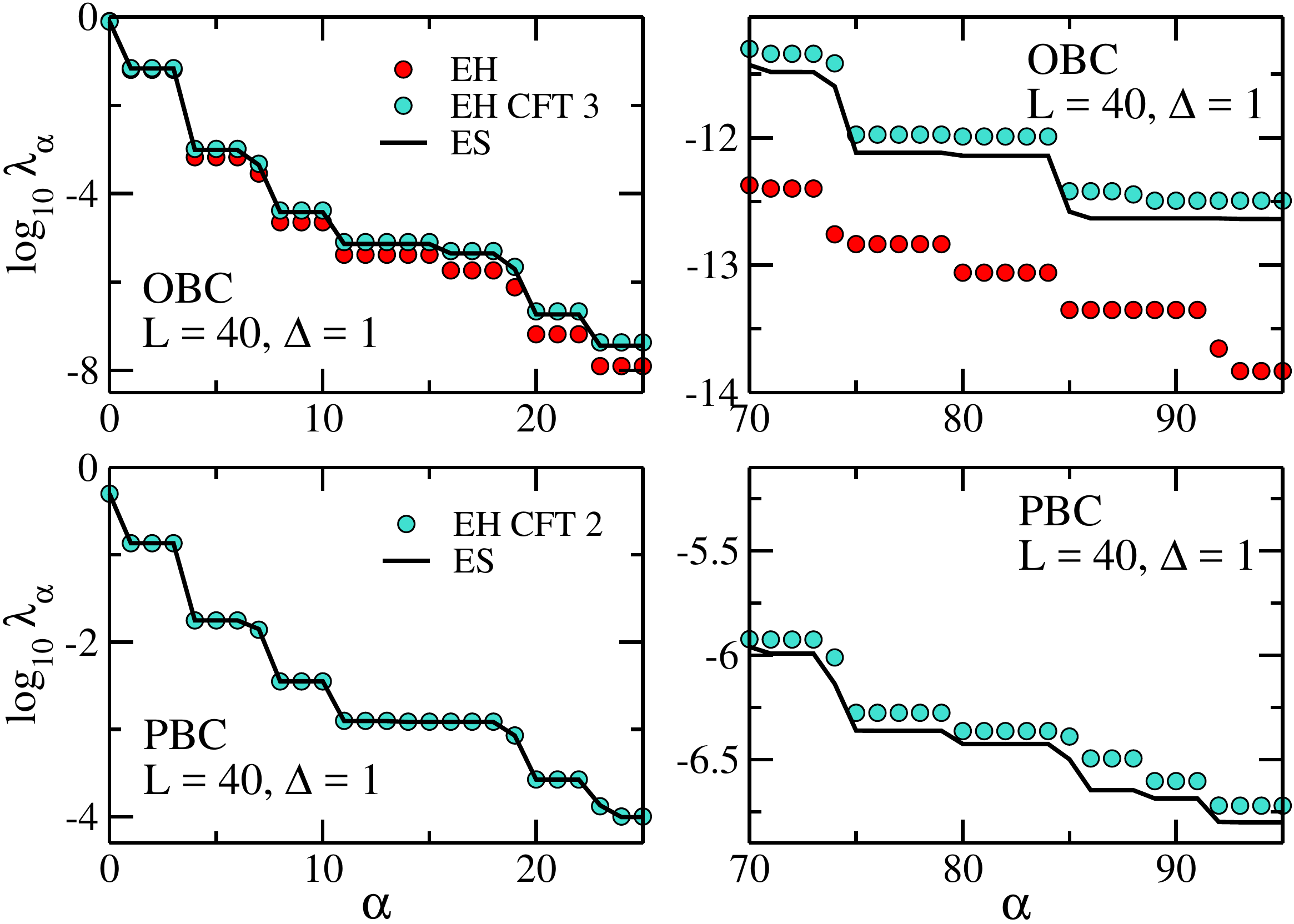}}}
\caption{Spectra comparison for the XXZ chain at the isotropic critical point. The black solid line is the exact ES. Red circles are the $\rho_{EH}$ eigenvalues computed via the infinite system EH (Eq.\eqref{BWtheorem} for OBC) and turquoise circles are computed via the CFT finite system EH (Eq.\eqref{BWtheoremCFT3} for OBC and Eq.\,\eqref{BWtheoremCFT2} for PBC, respectively). On the left $\lambda_\alpha$s with $\alpha=0,\dots 25$ and on the right with $\alpha=70,\dots,95$.} 
\label{xxz_fig_spec2} 
\end{figure}

\begin{figure}[]
{\centering\resizebox*{8.7cm}{!}{\includegraphics*{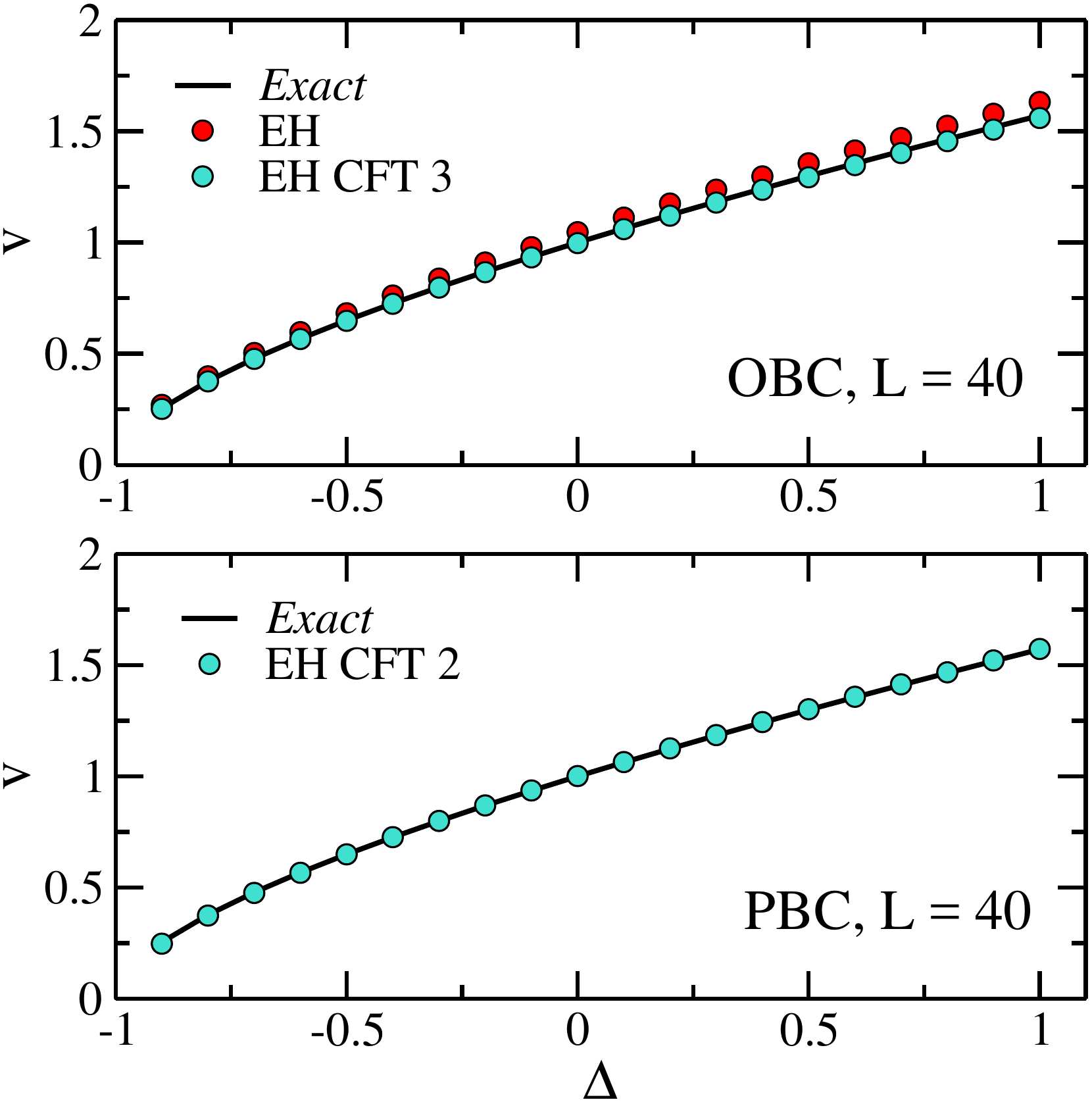}}}
\caption{Sound velocity exctracted via Eq.\,\eqref{vfromEH} from the first two eigevalues in the exact ES of a finite chain and in the EH spectrum, by using both the infinite system EHs (red circles) (Eq.\eqref{BWtheorem} for OBC) and the finite system EHs (turquoise circles) (Eq.\eqref{BWtheoremCFT3} for OBC and Eq.\,\eqref{BWtheoremCFT2} for PBC). The result is plotted against the exact expression Eq.\,\eqref{vxxz} for the sound velocity as a function of the anisotropy parameter $\Delta$.}
\label{xxz_fig_vel} 
\end{figure}

As a by-product of BW theorem we can use the exact ES and the EH spectrum to compute the sound velocity of the model. Indeed the relation between the two sets of eigenvalues reads $\lambda_\alpha = \exp{ ( -2 \pi / v  \,\, \varepsilon_\alpha ) }/Z$. 
We can take the ratio of two $\lambda$s to eliminate the normalization constant and we can invert this relation to get
\be \label{vfromEH}
v_\alpha = \frac{ 2 \pi ( \epsilon_\alpha - \epsilon_0 ) }{ \log \lambda_0/ \lambda_\alpha }.
\ee 
The result should be independent of $\alpha$ and this is indeed the case within negligible relative error. 
In Fig.\,\ref{xxz_fig_vel} we plot the sound velocity for $\alpha=1$ as a function of $\Delta$ against the exact result \eqref{vxxz}.
For OBC,  we also use the infinite system EH \eqref{BWtheorem} finding deviations only of few percent, as evident from the figure.

\begin{figure}[]
{\centering\resizebox*{8.7cm}{!}{\includegraphics*{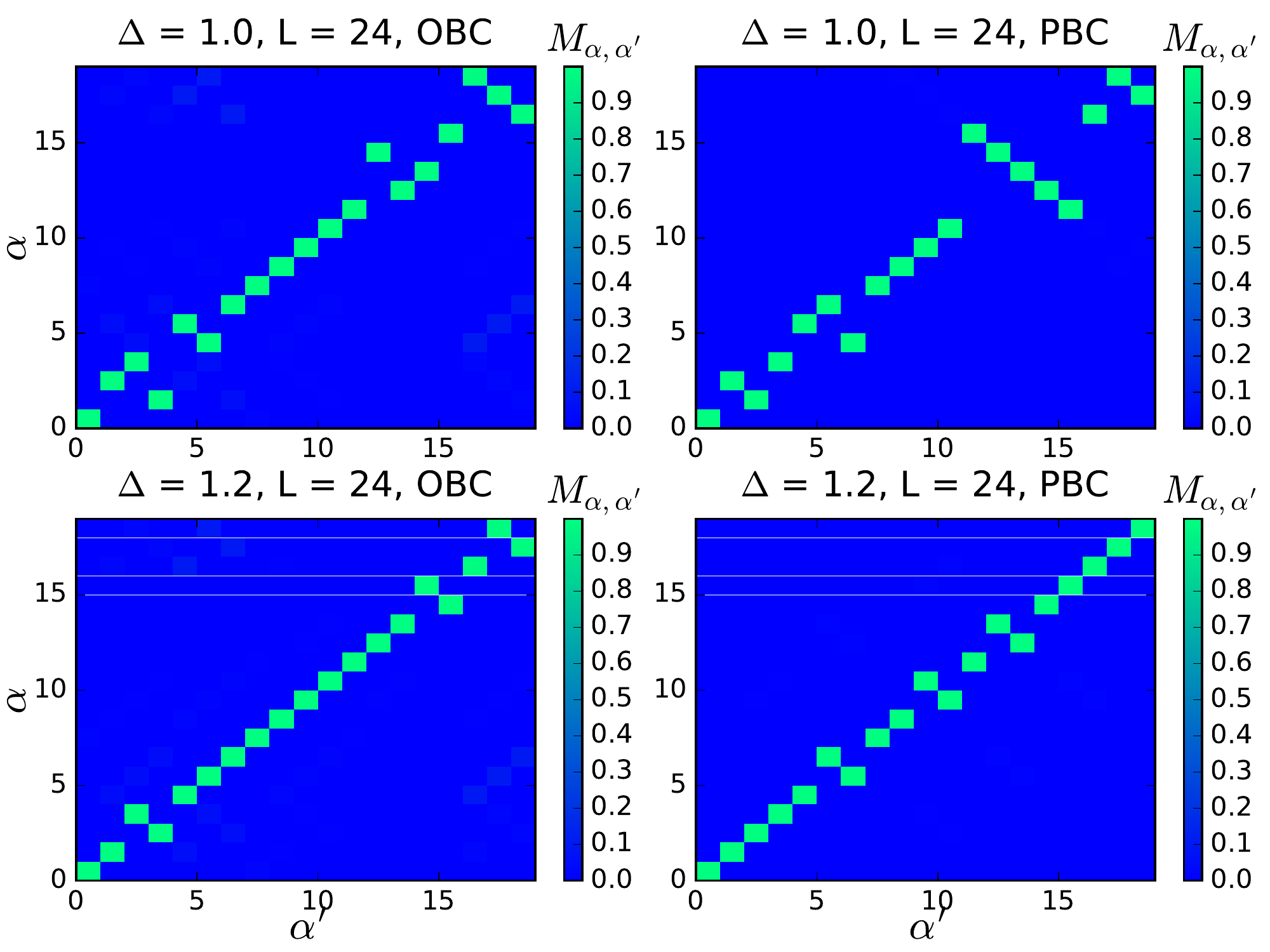}}}
\caption{Overlaps as defined in Eq.\,\eqref{overlaps} for the XXZ spin-$1/2$ chain at the BKT point $\Delta=1.0$ and in the N\'eel phase $\Delta=1.2$. The few points close to the diagonal correspond to exact degeneracies in spectrum.} 
\label{xxz_fig_ov} 
\end{figure}

Overlaps between $\rho_A$ and $\rho_{EH}$ eigenvectors are depicted in Fig.\,\ref{xxz_fig_ov} for $\Delta = 1$ and $\Delta = 1.2$. The two sets of eigenvectors match almost perfectly both in the OBC and PBC cases, as in the other models considered so far. 
This time we do not observe off-diagonal non-vanishing overlaps even when the CFT EH Eq.\,\eqref{BWtheoremCFT2} is used in the gapped phase. We attribute that to the fact that inverse correlation length, while being finite, is still of the order of the system size studied.

\begin{figure}[]
{\centering\resizebox*{8.7cm}{!}{\includegraphics*{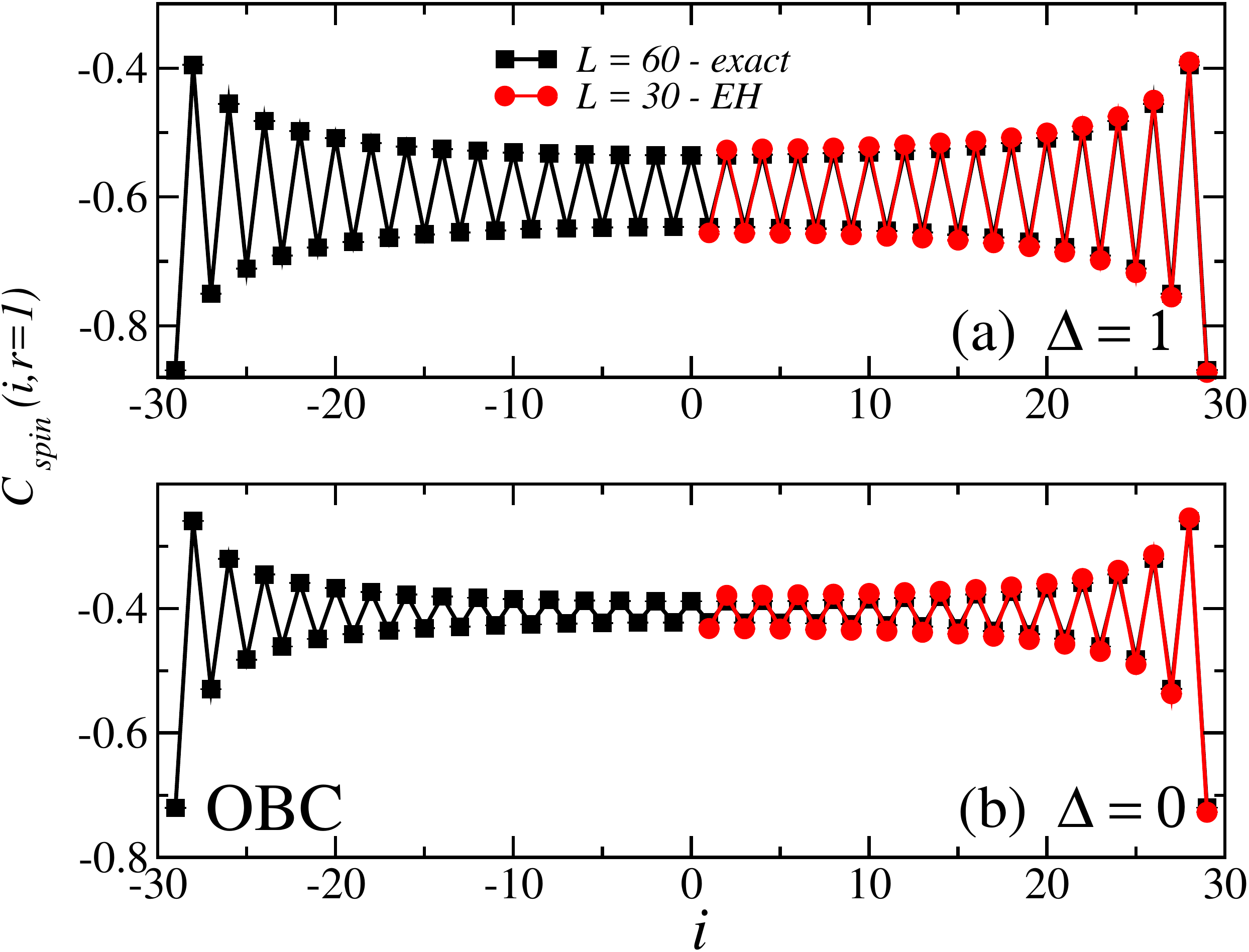}}}
\caption{Nearest-neighbour spin correlation function \eqref{Cspin} as function of position $i$ for chains with OBC for (a) $\Delta = 1$ and (b) $\Delta = 0$.
The square (black) and circle (red) points are respectively the exact results and those from EH-BW.
The Friedel oscillations are perfectly described by the EH approach.} 
\label{fig1} 
\end{figure}

\begin{figure}[]
{\centering\resizebox*{8.7cm}{!}{\includegraphics*{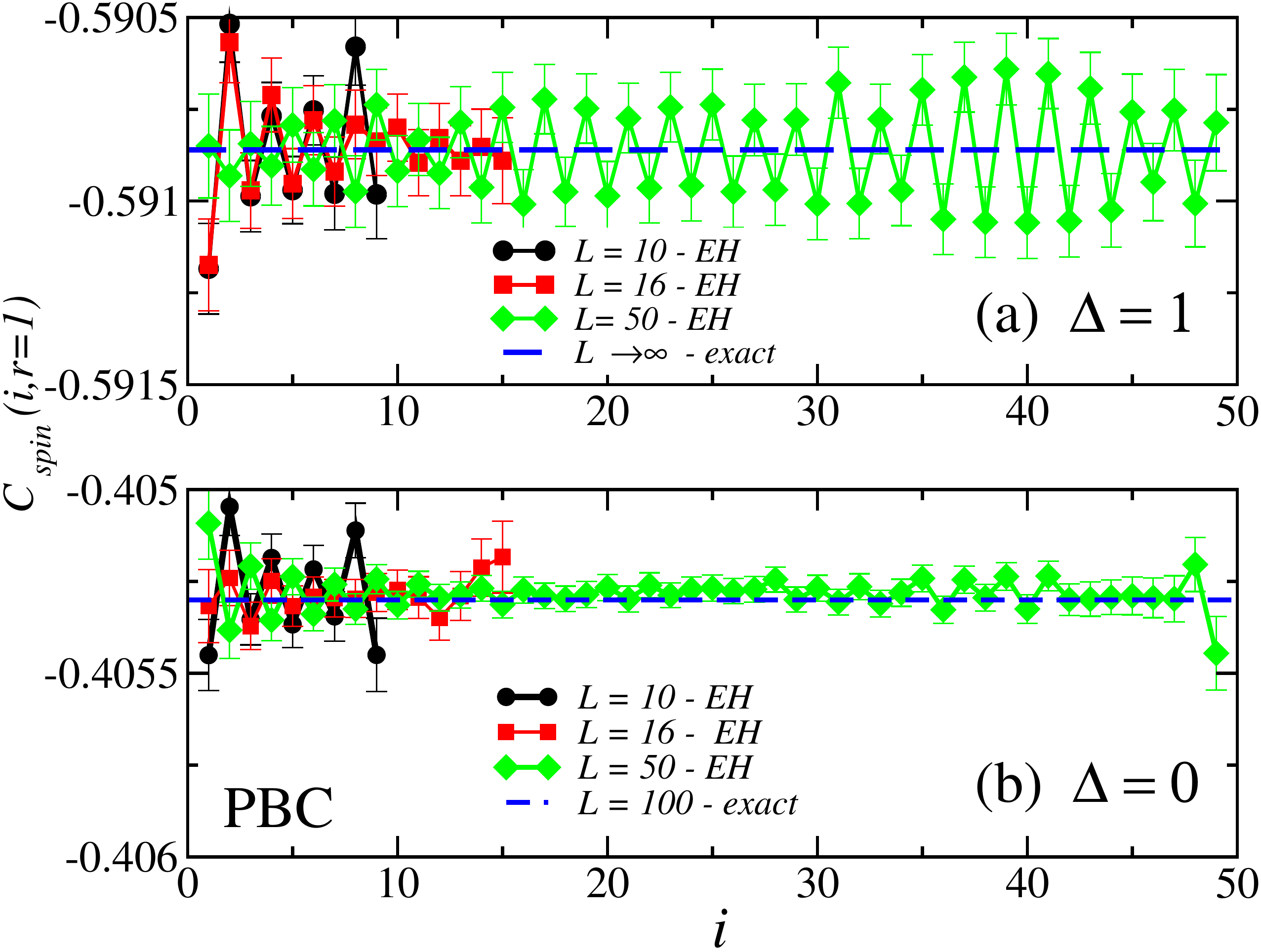}}}
\caption{Same as Fig.\,\ref{fig1} for chains with PBC.
The different points are results from the EH-BW with different sizes $L$. 
The horizontal lines are the exact results.
The relative deviations of the homogeneous result for $C_{spin}(i,r=1)$ is less then $0.1\%$. } 
\label{fig2} 
\end{figure}

The two-point function we analyze for this model is the spin-spin correlation function
\begin{equation}
 C_{spin}(i, r) = \left< S_{i}^z S_{i+r}^z\right>,
 \label{Cspin}
\end{equation}
that we compute using QMC. 
In this section we want to probe the thermodynamic values of this correlation. 
For this reason we do not exploit the finite size formulas for BW EH, but the infinite size ones \eqref{BWtheorem} and \eqref{BWtheoremCFT} 
which we apply to OBC and PBC respectively.
The results for $r=1$ are reported in Fig.\,\ref{fig1} and Fig.\,\ref{fig2} for the two cases, respectively. 
The two point of the phase diagram considered are the XX free fermions point $\Delta = 0$ and the BKT point $\Delta=1$. The velocity in the entanglement temperature is provided by Eq.\,\eqref{vxxz}. 
In the OBC case the agreement is perfect also close to the boundary where the system has been cut. 
The BW EH reproduces very well also the amplitude and the frequency of the Friedel oscillations caused by the free ends of the chain, with a relative error always smaller than 1\%. 

In the PBC case the ground state average is homogeneous. In fact the parabolic inhomogeneous coupling in Eq.\,\eqref{BWtheoremCFT} suppresses the boundary effects which affect the non-translational invariant BW-EH. This is strongly reminiscent of sine-square deformation Hamiltonians, which are actually close in functional form to the LBW EH in the PBC case~\cite{Katsura_2012}. The result from the thermal average of LBW EH is indeed almost homogeneous and deviations from the expected ground state value 
are less than $0.1\%$.

\begin{figure}[]
{\centering\resizebox*{8.7cm}{!}{\includegraphics*{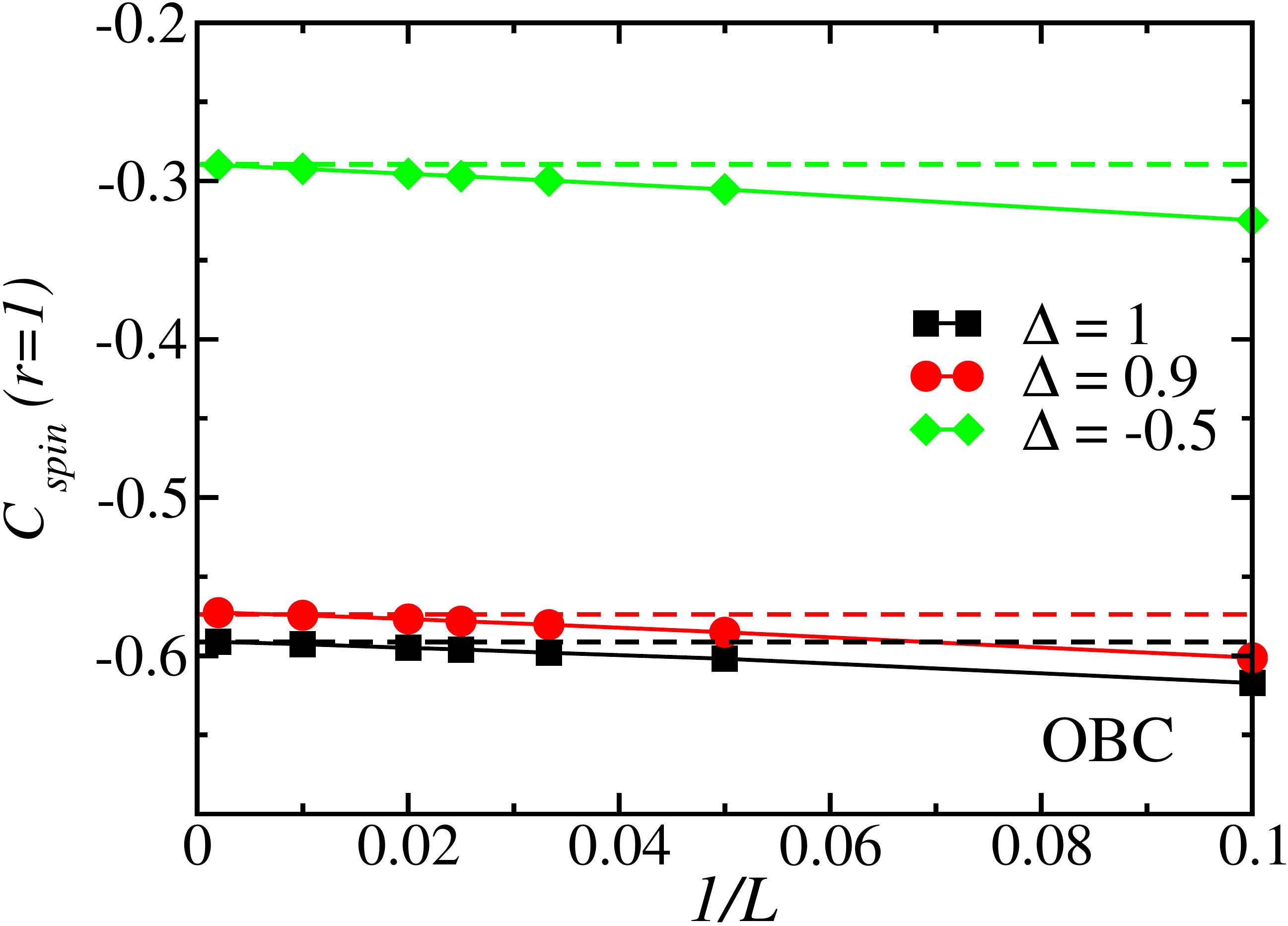}}}
\caption{Extrapolation of the average, $C_{spin}(r=1) = 1/L \sum_{i} C(i,r=1)$, to the thermodynamic limit, $L \to \infty$ 
for chains with OBC and different values of $\Delta$.
The horizontal lines represents the exact values of $C_{spin}(r=1)$ for $L \to \infty$.
In all the cases the EH-BW results converge to the exact ones in the limit $L \to \infty$.} 
\label{fig3}
\end{figure}

\begin{figure}[]
{\centering\resizebox*{8.7cm}{!}{\includegraphics*{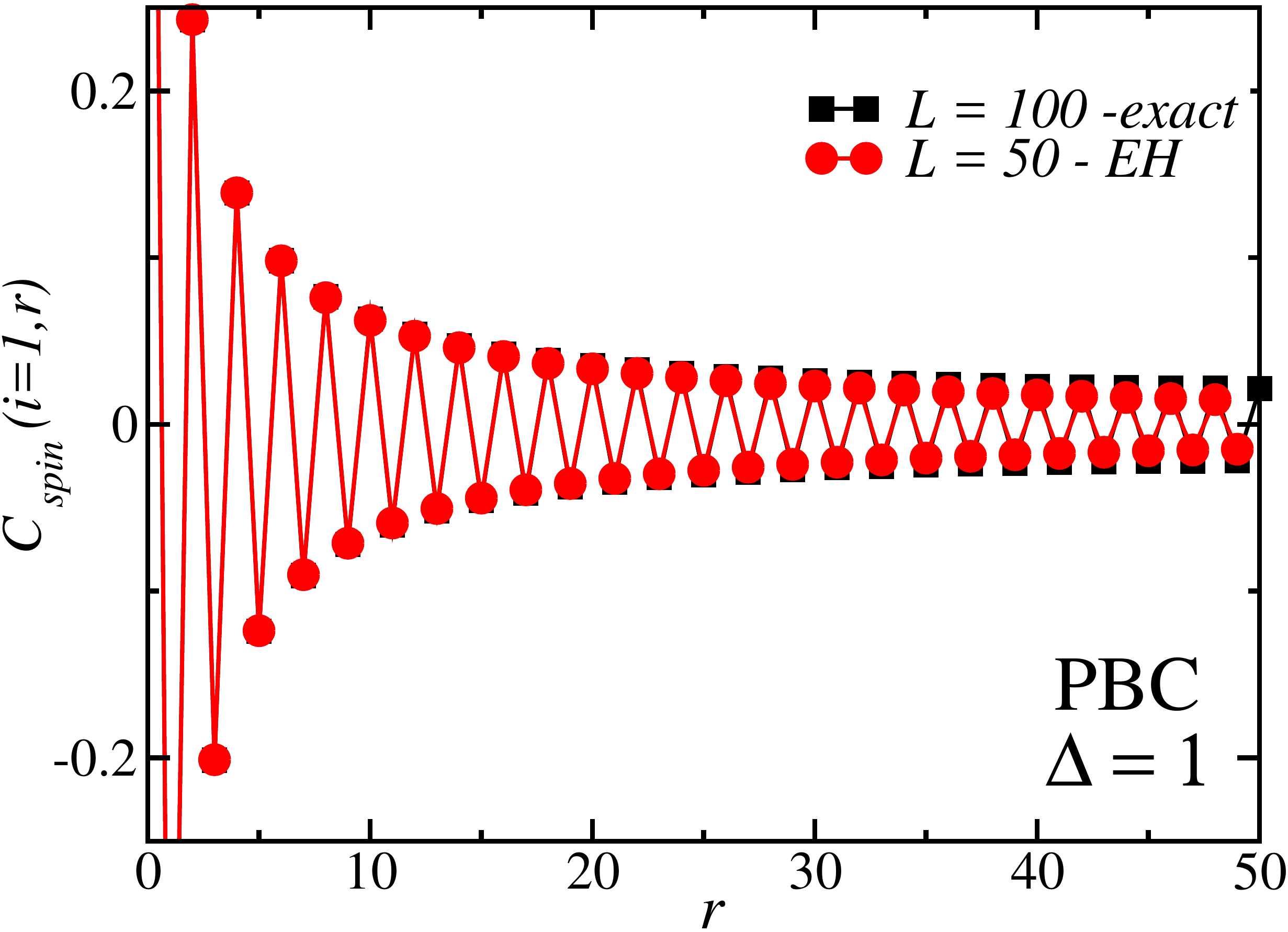}}}
\caption{Decay of spin-spin correlations for the isotropic case, $\Delta = 1$. 
We compare the exact data (black squares) with the EH results (red circles). } 
\label{fig4} 
\end{figure}

Fig.\,\ref{fig3} shows the finite size scaling of the LBW-EH $C_{spin}(i, 1)$ with OBC, averaged over the whole chain, against the exact TD limit value (dashed line) for three values of the anisotropy parameter $\Delta=-0.5,0.9,1.0$. The result strongly indicates that the field theory prediction of BW theorem Eq.\,\eqref{BWtheorem} is exact when $L \to \infty$. 
In Fig.\,\ref{fig4} we analyse the separation dependence of the 2-point function \eqref{Cspin}  at the isotropic point $\Delta=1$. 
Small deviations (of order 1\% on average with respect to $r$) are observed when $r$ becomes of the order of the correlation length, as expected.

\subsection{$S=1$ XXZ Model}
The isotropic $S=1$ spin chain is the archetypical model of a symmetry protected topological phase (SPTP)~\cite{haldane1983nonlinear,Regnault:2015aa,fradkinbook}. This new state of matter is characterized by a gap in the bulk, an even-degenerate ES, zero-modes living at the ends of an open chain and carrying fractionalized quantum numbers~\cite{Li2008,Regnault:2015aa}. Moreover, long-range order associated with a hidden $\mathbb{Z}_2\otimes\mathbb{Z}_2$ symmetry  is captured by the non-local order parameters
\be \label{string_op}
C_{str}^{\alpha} ( i ,r )  = - \med{ S^\alpha_{i}  \prod_{j= i + 1 }^{ i + r -1 } \exp \( i \pi S^\alpha_j \)   \, S^\alpha_{i+r}   },
\ee
which is non-vanishing for $r \to \infty$ for $\alpha = x,y,z$ (we focus in the following on the $z$ component and drop the index $\alpha$). In this work we considered two points in the phase diagram of the XXZ spin-$1$ chain, whose Hamiltonian is the same as Eq.\,\eqref{xxz}, but with the $S^\alpha$ matrices being a the spin-1 representation of the rotation group. The Haldane SPT phase extends in the parameter region $0 < \Delta \lesssim 1.17$ and it is separated from a gapless XY phase (on the left) by a BKT phase transition at $\Delta = 0$ and from a N\'eel phase (on the right) by a second order $c=1/2$ phase transition at $\Delta \simeq 1.17$~\cite{Chen:aa}.

We computed the ratios Eq.\,\eqref{ratios} obtained from the ES and from the LBW-EH for the two values of the anisotropy parameter $\Delta = 0$ and $\Delta = 1$ (for additional results on the ES, see Ref.~\onlinecite{Dalmonte:2017aa}). 
In the former case, as already discussed in the previous sections, Eqs.\,\eqref{BWtheorem}, \eqref{eqdist2}, and \eqref{eqdist3} have to yield the proper EH 
for OBC, PBC, gapped and gapless systems. At $\Delta=1$ instead the gap spoils conformal invariance and we do not expect Eq.\,\eqref{eqdist2} to work accurately.

\begin{figure}[]
{\centering\resizebox*{8.7cm}{!}{\includegraphics*{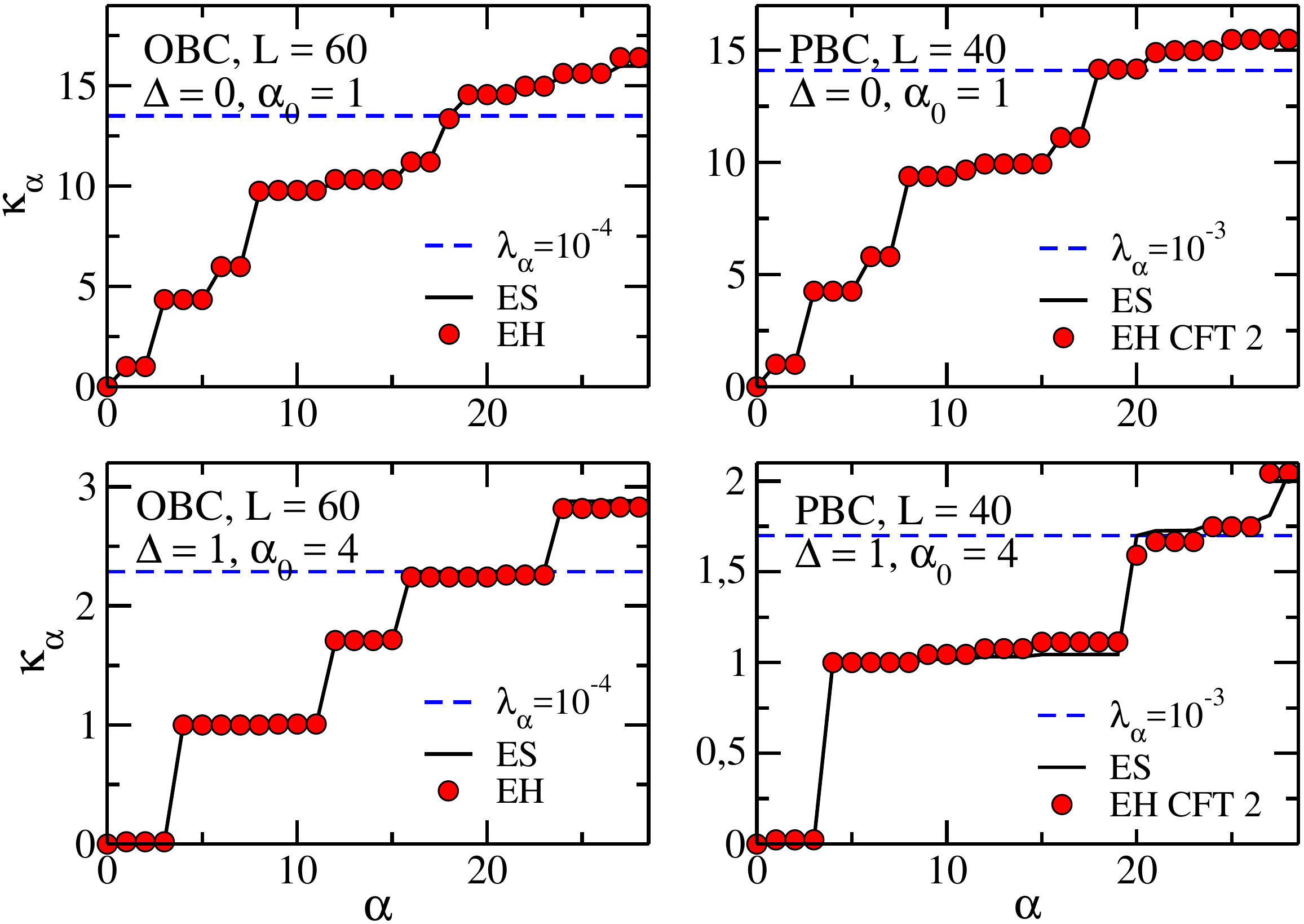}}}
\caption{Ratio $\kappa_\alpha$s for the XXZ spin $1$ chain. The black solid line and red circles stands for ratios computed from the exact ES and the EH spectrum respectively. The blue dashed line marks $\rho_A$ eigenvalues with a magnitude indicated in the legend. In the gapped Haldane phase $\Delta=1$ the EH Eq.\,\eqref{BWtheoremCFT} does not give the proper ES and relative discrepancies of the $\kappa_\alpha$s overcome 10\% for modestly large $\rho_A$ eigenvalues. } 
\label{ha_fig_spec}
\end{figure}

The results are plotted in Fig.\,\ref{ha_fig_spec}. At the $\Delta = 0$ BKT transition point maximum relative errors 
are 2\% for both the OBC and PBC case and for all the eigenvalues considered. In the gapped topological phase instead the same is true only when OBC are imposed on the system. In the PBC case relative discrepancies overcome 10\% (still, degeneracies are accurately reproduced). Note also that the typical ES even-degeneracy of the Haldane phase is perfectly captured by BW theorem in the OBC case. 

\begin{figure}[]
{\centering\resizebox*{8.7cm}{!}{\includegraphics*{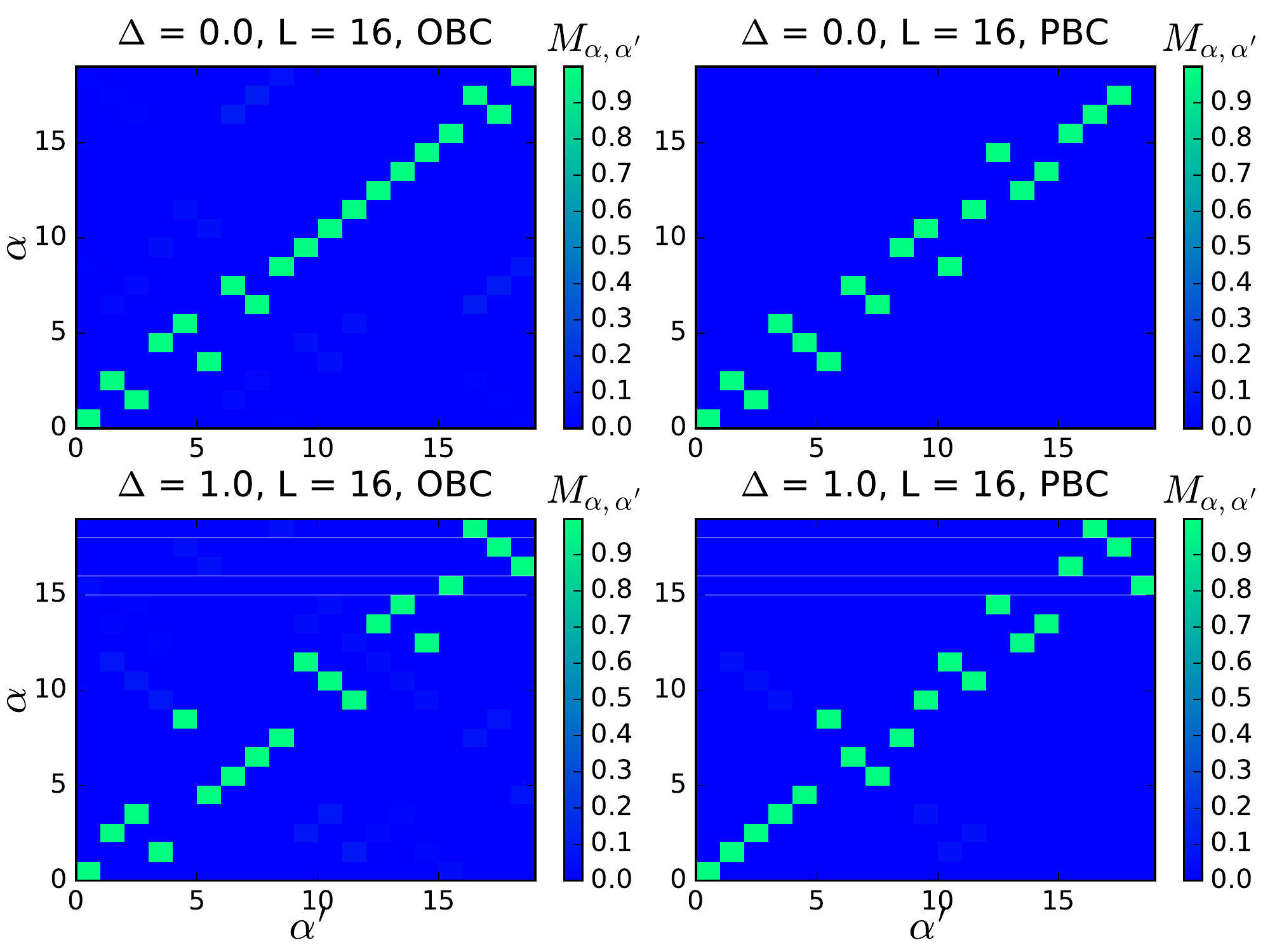}}}
\caption{Same as in Fig.\,\ref{is_fig_ov} for the $s=1$ XXZ chain at the BKT critical point $\Delta=0$ and in the Haldane phase $\Delta=1$. 
The several points outside the main diagonal all correspond to (almost) exact degeneracies in spectrum.} 
\label{ha_fig_ov} 
\end{figure}

Fig.\,\ref{ha_fig_ov} shows the density map of the overlaps defined in Eq.\,\eqref{overlaps} for the $\Delta = 0$ BKT critical point and for the $\Delta = 1$ isotropic point in the topological phase. Eigenvectors of $\rho_A$ and $\rho_{EH}$ overlap with deviations from unity of order $10^{-3}$ both in the gapped topological phase and at the gapless critical point. The reason why they are not properly ordered is the even (topological) exact degeneracy of the ES. The good agreement between the two sets of eigenvectors extends also when CFT EH Eq.\,\eqref{eqdist2} is used in the massive phase. This can be explained computing the correlation length of the system $\xi = v/m$. The gap at the isotropic point is known to be $m \simeq 0.40$,\cite{White1992} while the light velocity is estimated in the following to be $v \simeq 2.5$. Thus $\xi$ is larger than $6$ lattice spacings, making the CFT predictions very accurate despite the non-vanishing gap.\\

\begin{figure}[]
{\centering\resizebox*{8.7cm}{!}{\includegraphics*{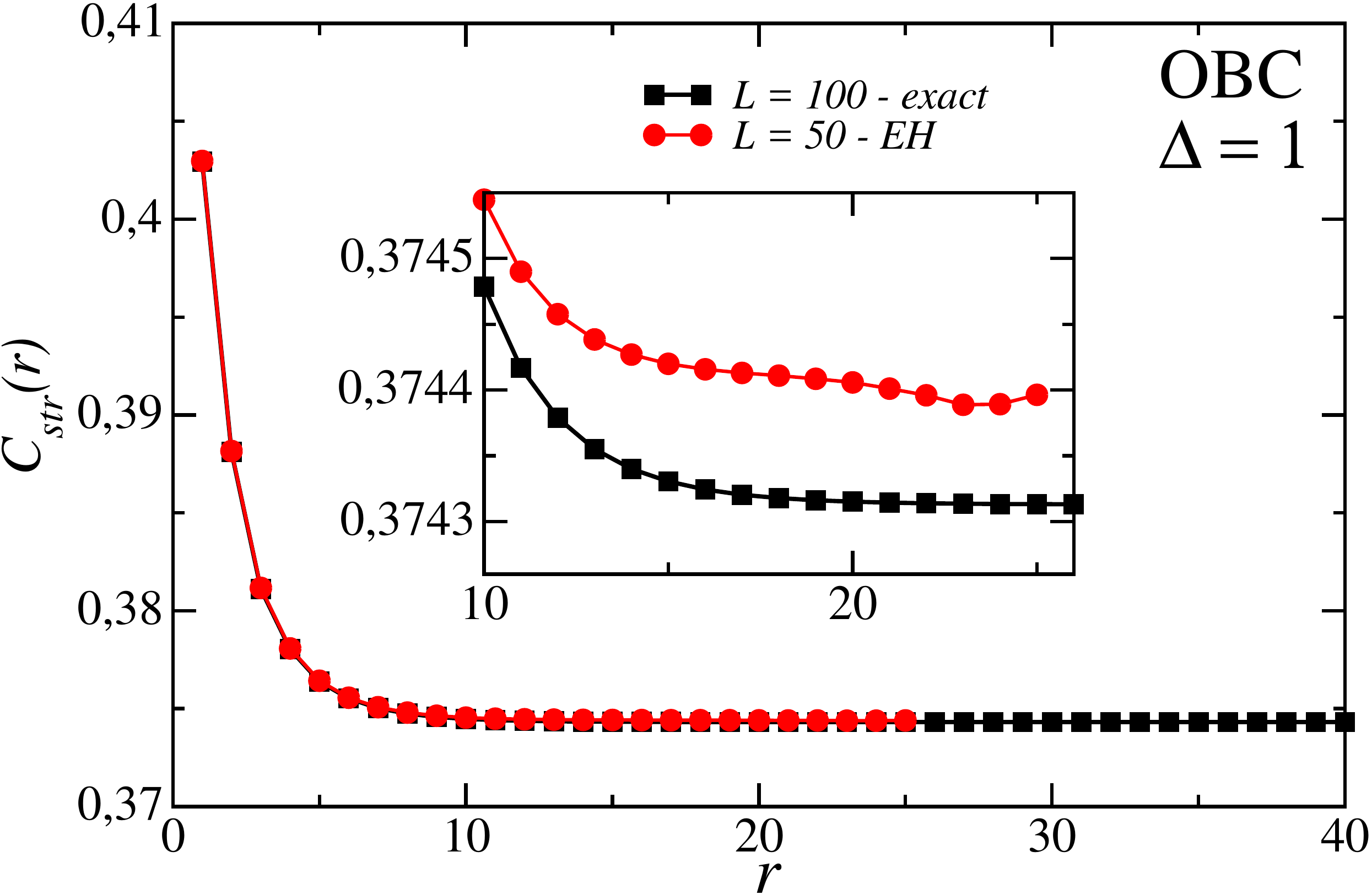}}}
\caption{Non-local order parameter as defined in Eq.\,\eqref{string_op} for the $s=1$ XXZ chain at $\Delta=1.0$.
The square (black) and circle (red) points are results for the original and the half-bipartion EH-BW systems, respectively. Relative errors are uniform and of order $10^{-4}$. } 
\label{ha_fig_corr} 
\end{figure}

The observable we test in this model is the non-local order parameter \eqref{string_op} along $z$ in the Haldane phase at $\Delta = 1$. 
In this case the light velocity necessary to compute the entanglement temperature is not known. 
We thus compute it using the relation between $\rho_A$ and EH eigenvalues as discussed in \ref{xxzsec} by using Eq.\,\eqref{vfromEH}. The result we get is independent of $\alpha$ within few percent relative error, both in the OBC and PBC cases and for all the lowest 30 eigenvalues computed. 
We then tuned $\beta$ in order to remove completely boundary effects close to the cut. In this way we get a sound velocity $v = 2 \pi/\beta = 2.475$. 
The results for the string order parameter are reported in Fig.\,\ref{ha_fig_corr}. The data correspond to a string starting in the middle of the right half-subsystem ($i=L/4$) 
and ending in the middle of the left half ($i=-L/4$).  
Relative deviations of the ground state average from the thermal expectation value are uniformly of order $10^{-4}$.

\subsection{$J_1-J_2$ Model}
As a final test for the 1-dimensional case, we discuss the $J_1-J_2$ quantum model, which includes next-to-nearest-neighbour interactions. The Hamiltonian of this spin-$1/2$ quantum chain reads:
\be 
H= J_1 \sum_i \vec{S}_i \cdot \vec{S}_{i+1} + J_2 \sum_i   \, \vec{S}_i \cdot \vec{S}_{i+2} ,
\ee
where $J_1,J_2>0$ will be considered in what follows.
When $J_2=0$ this model coincides with the XXZ chain at the BKT critical point. When $J_2$ is switched on, the model remains critical for a finite interval in $J_2/J_1$ and it is described by a $c=1$ CFT with the same Luttinger parameter $K=1/2$ throughout the whole interval\cite{haldane1982}. When $J_2/J_1$ reaches the approximate value $J_2/J_1 \simeq 0.2411$ \cite{okamoto1992}, a gap opens and the system enters a dimerized phase characterized by a non vanishing dimer-order parameter
\be 
d_i = \med{ \vec{S}_{2 i -1} \cdot \vec{S}_{2 i} } - \med{ \vec{S}_{2 i} \cdot \vec{S}_{2 i+1} }.
\ee
This phase contains an exactly solvable point for $J_2=J_1/2$ \cite{mg1969}, where the ground state factorizes into a product of spin $1$ singlets on adjacent sites:
\be \nonumber
\ket{\psi} = \bigotimes_{a=1}^{L/2} \frac{ \ket{ \uparrow \, \downarrow \, } - \ket{ \downarrow \, \uparrow \, } }{ \sqrt{2} }.
\ee
At this fine-tuned point the entanglement spectrum of the system is trivial, with either one or two equal non-vanishing eigenvalues (depending on the cut). The same is not true for the BW-EH spectrum. Moreover the gap in the dimerized phase is maximum when $J_2 \simeq 0.6 J_1$ \cite{white1996} and it slowly decreases with increasing $J_2$. We thus expect the worst results to be observed after the Majumdar-Ghosh factorization point with finite $J_2$. Note that the same behavior is expected close to any factorizable point, as discussed in the context of the Affleck-Kennedy-Lieb-Tasaki spin-1 chain in Ref.~\onlinecite{Dalmonte:2017aa}. 
When $J_2 \to \infty$ the system reduces to  two independent critical Heisenberg chains.

\begin{figure}[]
{\centering\resizebox*{8.7cm}{!}{\includegraphics*{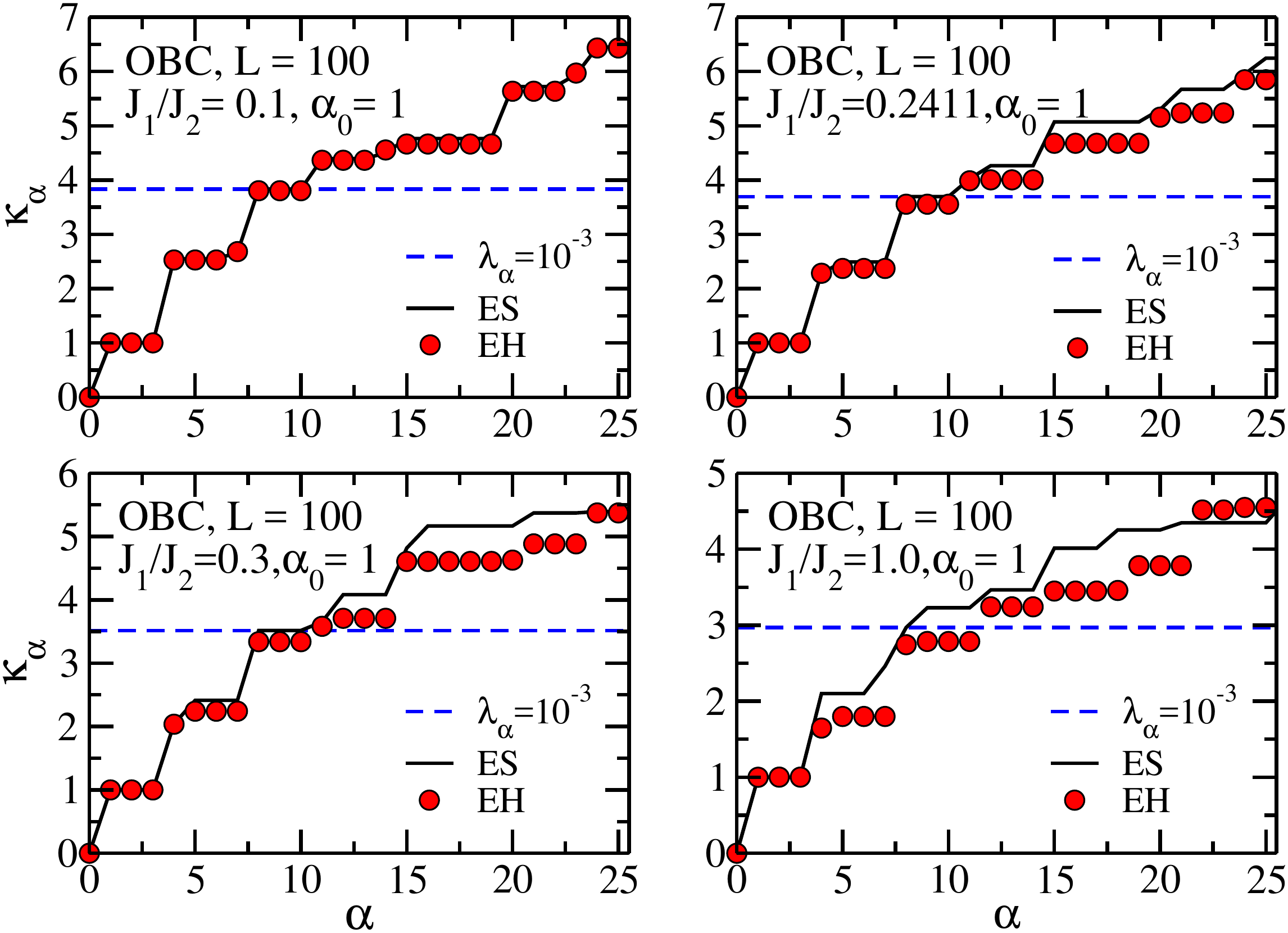}}}
\caption{Ratio $\kappa_\alpha$s for the $J_1-J_2$ chain with OBC. The black solid line and red circles stands for ratios computed from the exact ES and the EH spectrum respectively. The blue dashed line marks $\rho_A$ eigenvalues with a magnitude indicated in the legend. Agreement with BW theorem is good in the middle of the critical phase and close to it (relative deviations smaller than 10\%), while it gets worse as $J_2/J_1$ is increased away from the critical line (relative deviations larger than 20\%). } 
\label{mg_fig_spec} 
\end{figure}

Fig.\,\ref{mg_fig_spec} shows the universal ratios Eq.\,\eqref{ratios} comparison between entanglement spectrum and LBW-EH spectrum for the OBC case where the system is in the middle of the critical phase ($J_2 = 0.1 J_1$), at the critical point ($J_2=0.2411 J_1$) and in the dimerized phase ($J_2 = 0.3 J_1$ and $J_2 = J_1$). In the first two cases the largest relative deviations are 2\% and 8\% respectively. When $J_2 = 0.3 J_1$ the gap is small and large discrepancies affect only eigenvalues in the ES smaller than $10^{-3}$, where they reach 10\% relative error. When $J_2 = J_1$ the gap is much larger and BW theorem does not reproduce the correct ratios for eigenvalues of the $\rho_A$ of magnitude $10^{-2}$ and their degeneracies. Relative deviations are larger than 20\% also for the first 10 ratios.\\

\begin{figure}[]
{\centering\resizebox*{8.7cm}{!}{\includegraphics*{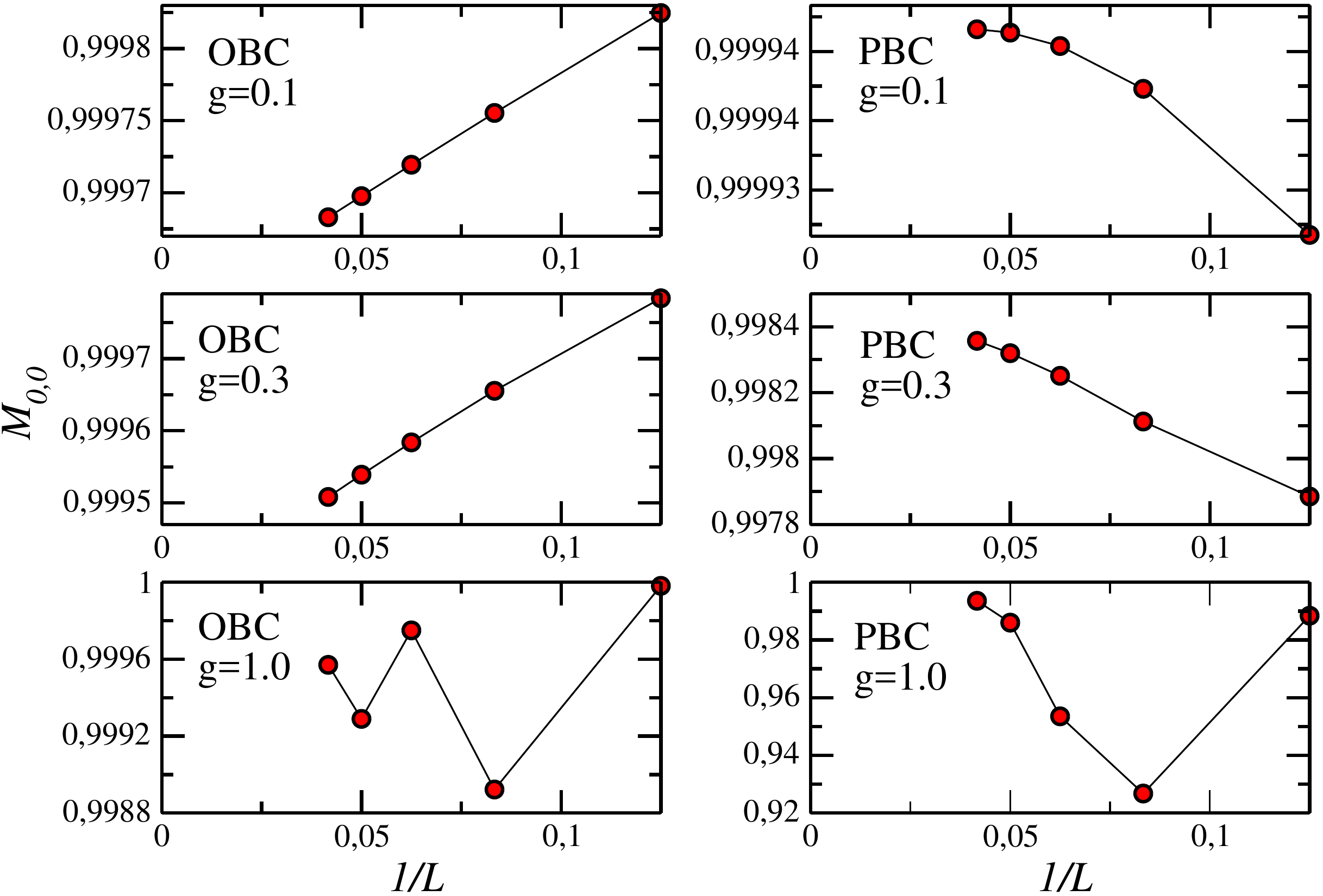}}}
\caption{Finite size scaling of the ground state overlap $M_{0,0}$ as defined in eq. Eq.\,\eqref{overlaps} for the $J_1-J_2$ chain for both OBC and PBC and for the three distinct values of the coupling constant $g=0.1$ (gapless critical phase), $g=0.3$ (gapped phase close to the critical point), $g=1.0$ (gapped phase with large gap). In the OBC case the overlap decreases with the systems size, while it increases towards unity in the PBC case.}
\label{mg_fig_eigv} 
\end{figure}

In Fig.\,\ref{mg_fig_eigv} we also show the finite size scaling of the overlaps Eq.\,\eqref{overlaps} for the ground state. The overlap always increases for large system sizes in the PBC case, while it is decreasing in the OBC case for the lengths accessible via ED. We attribute the non monotonic behaviour at $g=1$ as a signal of the dimer order in the gapped phase.

\section{Two-dimensions}
\label{sec_2d}

Differently from the one-dimensional case discussed so far, direct studies of entanglement Hamiltonians in interacting 2D and 3D models are lacking apart from 
few aforementioned cases discussed within perturbation theory.
As such, the potential of applying the BW theorem reliably to lattice model can be of even stronger impact than in 1D systems.
A closely related subject concerns the topological matter, where Li and Haldane conjectured that the low-lying part of the ES is capturing the edge mode energetics~\cite{Li2008}. 

In this section, we check the applicability of the LBW EH for the two-dimensional XXZ model on a square lattice, defined as:
\begin{equation} 
\label{xxz2d}
H= J \sum_{\left<i,j\right>} \( S^x_i S^x_j + S^y_i S^y_j + \Delta S^z_{i} S^z_j   \),
\end{equation}
for both the cylinder and the torus  geometries for which we employ Eq. \eqref{eq_dist} and \eqref{eq_dist1} respectively.

We focus in two cases: (i) the isotropic, $\Delta =1$, and (ii) the XX, $\Delta = 0$ points.
For these two values of $\Delta$, the ground state of the system spontaneously breaks the continuous SU($2$) and U($1$) symmetries, respectively \cite{sandvik1997,sandvik1999}. 
In the first case, the low-lying field theory is a $\mathbb{CP}(1)$ model.
In both cases, the low energy degrees of freedom of the system are characterized by a linear dispersion relation, and Lorentz-invariance emerges as an effective low-energy symmetry.

Differently from the one-dimensional cases considered in the last section, exact diagonalization approaches are of little use here, as the LBW-EH approach cannot exploit lattice symmetries, as is limited to very small lattices where universality is most probably spoiled by finite volume effects. 

As such, we do not attempt comparisons based on entanglement spectrum
and eigenvectors of BW-EH, but rather focus directly on the expectation values of first-neighbour correlation function
\begin{equation}
 C_{spin}(\vec{i},\vec{r}) = \left< S_{\vec i}^z S_{\vec i+\vec r}^z\right>,
 \label{Cspin2d}
\end{equation}
and the AFM order parameter. 
The sound velocities $v = 1.657J$ ($\Delta = 1$)  and $v=1.134J$ ($\Delta = 0$) obtained in Refs.~\onlinecite{wiese1998} and  \onlinecite{sandvik1999}, respectively, are used to calculate  $\beta_{EH} = 2\pi/v$.

\begin{figure}[]
{\centering\resizebox*{8.7cm}{!}{\includegraphics*{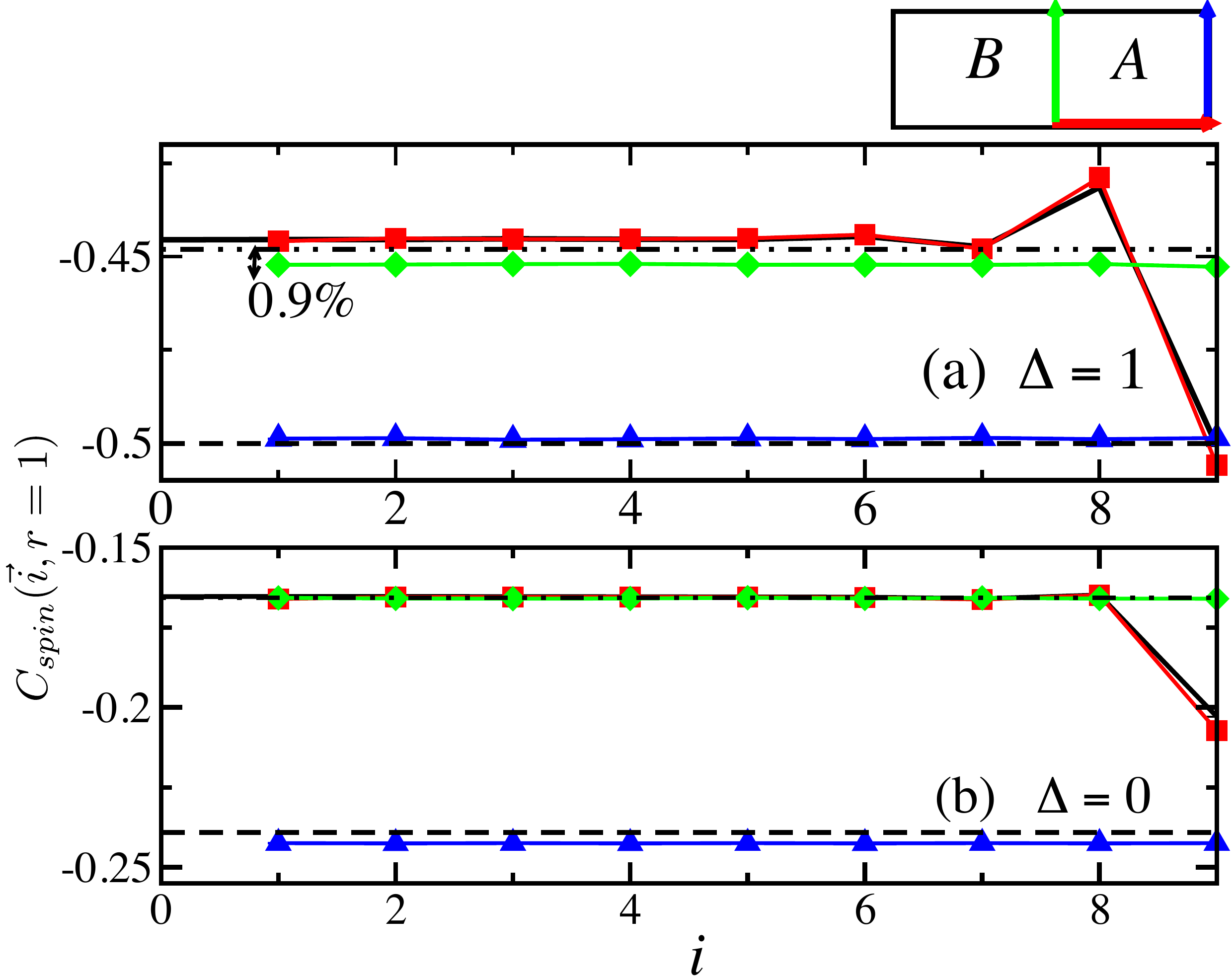}}}
\caption{Nearest-neighbor spin correlation function as function of bond index $i$ for the 2D square lattice in the cylinder (BW) geometry for (a) $\Delta =1$ and (b) $\Delta = 0$.
The different points are the results of $C_{spin}(\vec{i},r=1)$ in  different paths of the 2D system: red squares, blue triangles and green diamonds are along  
$(i_x=i,i_y=1)$, $(i_x=L,i_y=i)$ and $(i_x=1,i_y=i)$, respectively; see cartoon. 
In addition, the black curves are the exact ground state results for the system  $20 \times 10$.} 
\label{Cfn2dOBC} 
\end{figure}

\begin{figure}[]
{\centering\resizebox*{8.7cm}{!}{\includegraphics*{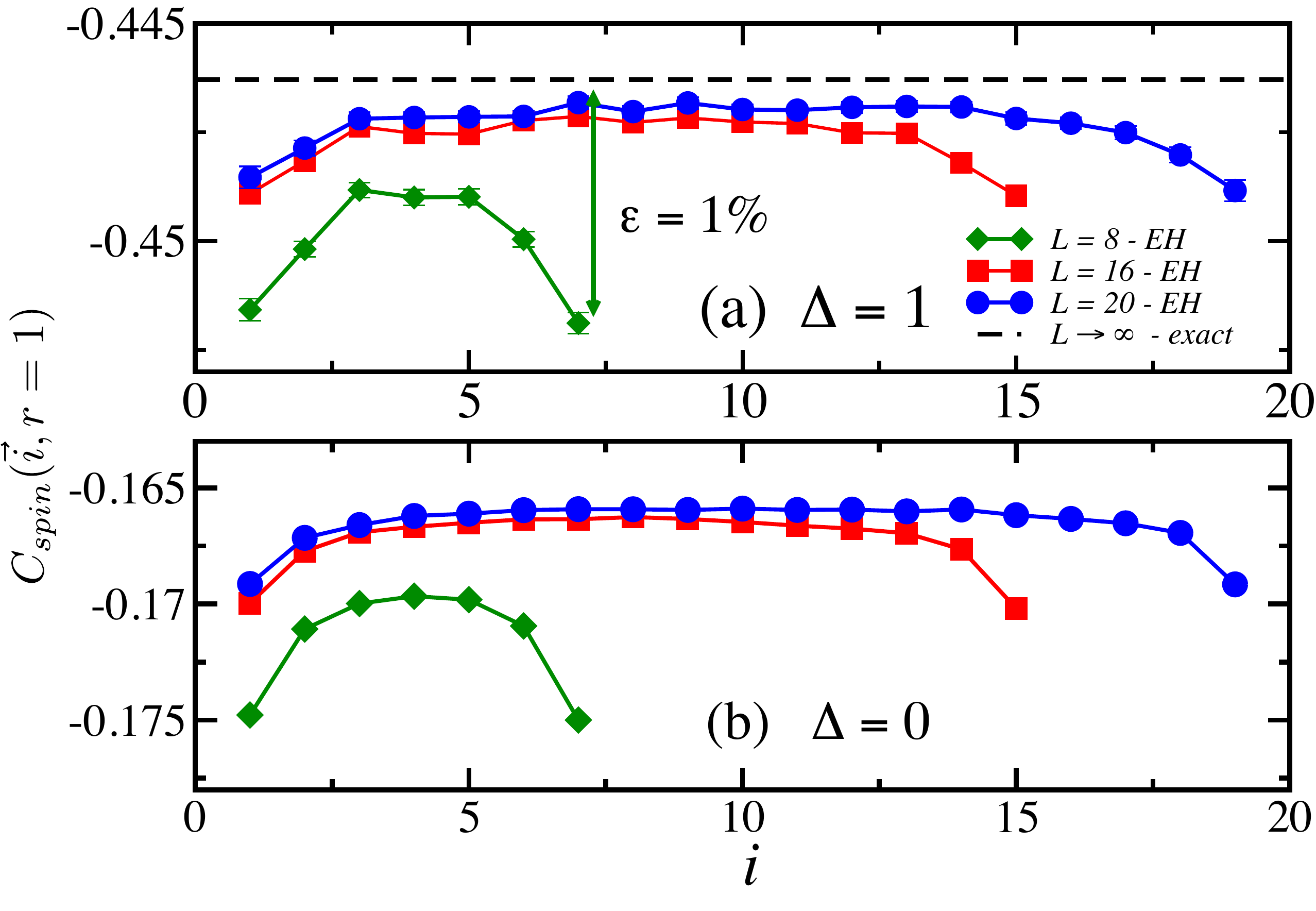}}}
\caption{ $C_{spin}(\vec{i},r=1)$ as function of bond index $\vec{i} = (i_x = i, i_y=1)$ for different system sizes, $L$, in the 2D torus (CFT1) geometry.
(a) $\Delta =1$ and (b) $\Delta = 0$.
The dashed horizontal line for $\Delta = 1$ is the exact result of $C_{spin}(\vec{i},r=1)$ extrapolated to the $L \to \infty$ limit \cite{sandvik1997}.} 
\label{Cfn2dPBC} 
\end{figure}

\begin{figure}[]
{\centering\resizebox*{8.7cm}{!}{\includegraphics*{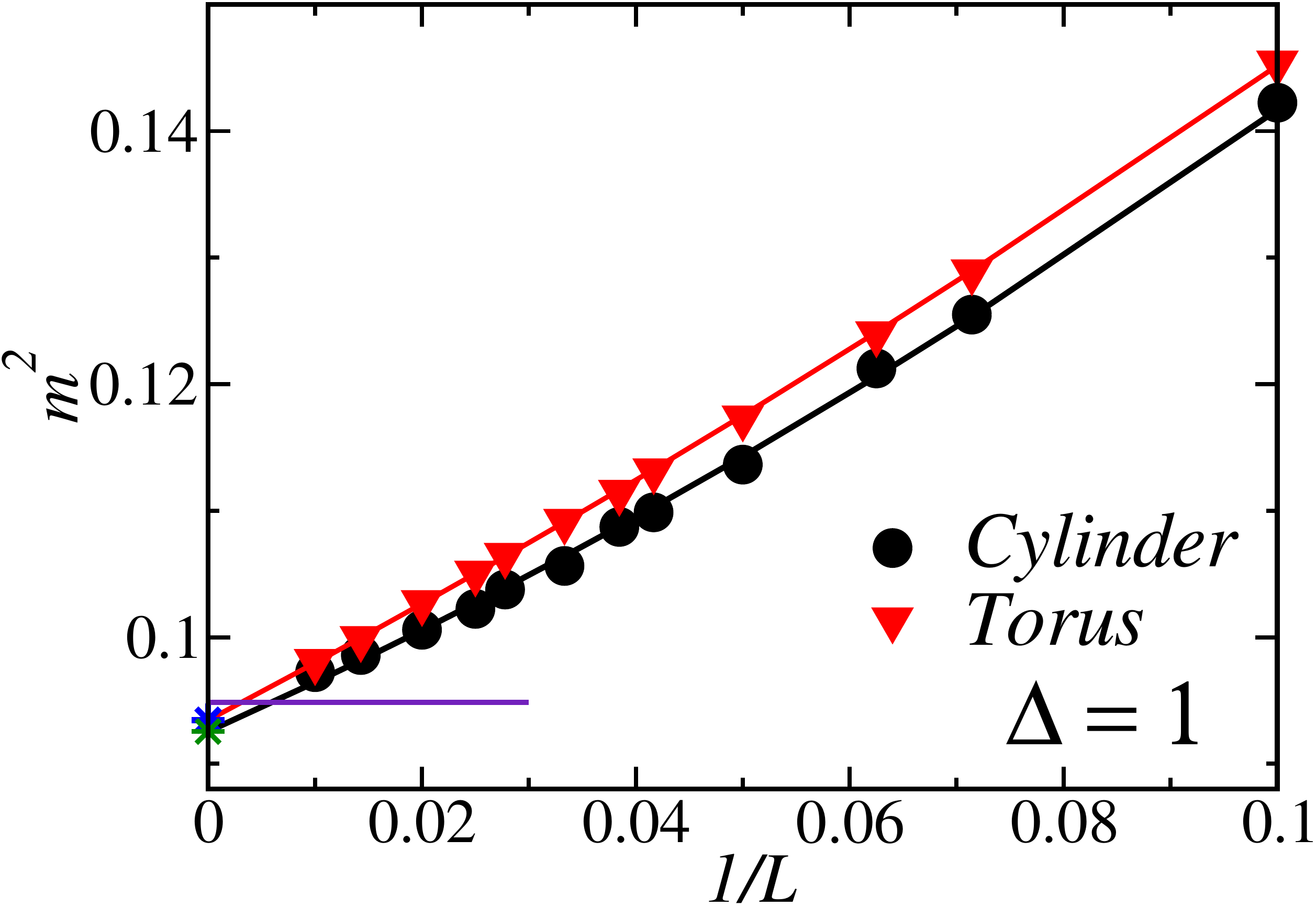}}}
\caption{Finite-size scaling of $m^2$ for the EH-BW system ($L \times L$) in both the cylinder (black-square) and torus (red-circle) geometries. 
The horizontal line represent the exact QMC value of $m^2$ obtained in Ref.~\onlinecite{wiese1998}.} 
\label{m2AFM} 
\end{figure}

First, we discus the comparison  of the BW-EH $C_{spin}(\vec{i},r=1)$ in the cylinder geometry with exact results, see Fig. \ref{Cfn2dOBC}.
Even for the relatively small system considered ($L = 10$), the agreement of  $C_{spin}(\vec{i},r=1)$ with the exact results is very good. The LBW-EH qualitatively reproduces the behaviour of $C_{spin}(\vec{i},r=1)$, and the relative errors are  $< 1\% $.
Larger deviations are observed for correlations along the boundary between the two subsystems, see green curve, and $\Delta = 1$ (these deviations are much milder in the anisotropic case).

In the toroidal geometry, the exact formula of the modular Hamiltonian is not known even in the continuum. 
Here, we heuristically employ Eq. \eqref{eq_dist1}. 
In Fig. \ref{Cfn2dPBC}, we plot the $C_{spin}(\vec{i},r=1)$ for different system sizes in the torus geometry. In this case, $C_{spin}(\vec{i},r=1)$ is almost homogeneous, with deviations smaller than $< 1\%$. 
Furthermore, as  $L$ increases the BW-EH results approaches the exact results obtained in the thermodynamic limit $L \to \infty$. This strongly suggests that the employed ansatz, whilst not necessarily exact, provides a very accurate description of 2D EH on tori.

Finally, we discuss if the BW-EH  describes the AFM long-range order in the $\Delta = 1$ case.
The AFM phase is characterized by the order parameter
\begin{equation}
 m^2 = \left< \frac{1}{N^2} \sum_{i,j} (-1)^{i+j} S_i S_j   \right>,
\end{equation}
where $i$ and $j$ run throughout all the sites of the system and $N$ is the total number of spins.
If AFM long-range order is present, $m^2$ is finite in the thermodynamic limit, since AFM correlations remain nonzero at large distances.
For a finite system, $2L \times L$, split in two equal halves of sizes $L \times L$,
one can write $m^2 = m_A^2 + m_B^2 + m_{A,B}^2 + m_{B,A}^2 $,
where $m_A^2$ and $m_B^2$ are the contributions of subsystem $A$ and $B$, respectively,
and $m_{A,B}^2$ and $m_{B,A}^2$ are contributions of crossing terms.
In the limit $N \to \infty$, all these four terms are equal, and $m^2 = 4m_A^2$.

Figure \ref{m2AFM} shows the finite-size scaling of $m^2$ obtained with the BW-EH.
As already observed for the first-neighbour correlation functions, $m^2$ is in  good agreement  with the exact result.
We obtain $m^2(L\to\infty) = 0.0925(4)$ and $m^2(L\to\infty) = 0.0934(1)$ for the  the cylinder and the torus geometries, respectively.
The relative errors with  QMC exact results,  $m^2(L\to\infty) = 0.0948$ \cite{wiese1998}, are $\epsilon_{m^2} \approx2.4\%$ and  $\epsilon_{m^2} \approx1.5\%$.

\section{Discussions and Conclusions}\label{sec_concl}

We have discussed an approach to systematically build approximate entanglement Hamiltonians of statistical mechanics models by applying the Bisognano-Wichmann theorem on the lattice. Starting from a recasting of the latter theorem on discrete space, we have presented a series of diagnostics based on the entanglement spectrum, the eigenvectors of the reduced density matrix, and expectation values of correlation functions. Based on these quantities, we have carried out numerical simulations for both 1D and 2D models whose low-energy physics is captured by a Lorentz invariant quantum field theory. 

In critical cases, such as conformally invariant points and phases in 1D, and spontaneous-symmetry-breaking phases in 2D, our results strongly support that the lattice Bisognano-Wichmann entanglement Hamiltonian captures very accurately all properties of the original system. 
What is particularly striking is that even the eigenvectors of the reduced density matrix have very large overlaps, which seem not to vanish with increasing system size. 
This last fact is particularly surprising, as overlaps are quantities that typically vanish in the TD limit, suggesting that there might be deeper connections between the structure of the EH and the BW theorem directly at the lattice level. 
Let us also remark that our results show that the modified CFT formulas obtained by Cardy and Tonni~\cite{Cardy:2016aa} cope extremely well with finite-lattice spacing
and, in fact, considerably reduce finite size effects when compared to the infinite-size BW EH.

In gapped systems, we typically find good agreement for both topologically trivial and non-trivial phases, with the exception of the $J_1-J_2$ model: in this last case, the effects of strong dimerization considerably spoil the applicability of field theory results, as the phase itself does not support a description in terms of smoothly varying fields. Somehow surprisingly, degeneracies of the ES are well captured, and even the overlap of the entanglement ground state is anomalously large. 

At the methodological level, our study shows that well-established statistical mechanics tools such as DMRG and quantum Monte Carlo can be applied without major effort to the investigation of entanglement Hamiltonians. A first potential application along this route is the potential to carry systematic entanglement spectroscopy with QMC, not relying on reconstructing the ES from R\'enyi entropies~\cite{Chung:2013aa}, but rather on monitoring correlation functions in the entanglement ground state, and extract the corresponding entanglement gaps from the decay of correlation functions. A second application concerns the possibility of further severely reducing finite-size effects when measuring correlation functions by directly accessing a finite bipartition of an infinite system~\footnote{This approach could also be employed in t-DMRG simulations.}. A third application is related to boosting procedures employed to extract the entanglement Hamiltonians given a ground state of interest, as discussed in two recent works~\cite{Toldin:2018aa,Zhu:2018aa}. Our general analysis supports from the theoretical side the results obtained for the models considered in these works. Furthermore, from the experimental side, our results immediately extend the regime of applicability of the approach proposed in Ref.~\onlinecite{Dalmonte:2017aa} to perform quantum simulation and spectroscopy of the EH, especially in two-dimensional interacting models.

The discussion we have presented here only concerns statistical mechanics models whose Hilbert space can be written in tensor product form. An open question is to extend this approach to lattice gauge theories: in this context, a lattice version of BW can be constructed using established methods to properly build reduced density matrices that consider the effect of Gauss law at the boundary between two partitions~\cite{Buividovich:2008aa,Casini:2014aa}. Another important feature of our approach is that, for critical systems, it is limited to quantum field theories with $z=1$. While this encompasses a very broad class of quantum critical points, it would be interesting to extend the Bisognano-Wichmann theorem beyond its original applicability regime, thus providing a direct link between the dispersion relation of equilibrium systems and their ground state entanglement properties. Extending this correspondence could shed further light into the origin of area law (and violations thereof) of entanglement in the ground state of lattice models~\cite{ecp-10}.

\acknowledgements{We acknowledge useful discussions with V. Alba, V. Eisler, M. Falconi, R. Fazio, N. Laflorencie, I. Peschel, S. Sachdev, E. Tonni, and K. van Acoleyen. MD thanks B. Vermersch and P. Zoller for previous collaboration on related work. 
This work is supported by the ERC under grants  numbers 758329 (AGEnTh) and 771536 (NEMO). 
TMS and MD acknowledge computing resources at Cineca Supercomputing Centre through the Italian SuperComputing Resource Allocation via the ISCRA grants TopoXY and QMCofEH. GG and TMS equally contributed to this work.}


\phantomsection
\addcontentsline{toc}{chapter}{Bibliography}
\bibliography{BWEHv2}

\appendix
\section{Autocorrelation times in Monte Carlo simulations of the lattice Bisognano-Wichmann entanglement Hamiltonian}

In this Appendix we discuss the efficiency of the Stochastic Series Expansion (SSE) method \cite{sandvik1991,sandvik2002}
in simulating  Bisognano-Whichmann entanglement Hamiltonians (BW-EH).
The BW-EH can be interpreted as an Hamiltonian with a local effective inverse temperature $\beta$ that increases away from the boundary.
Thus, far away from the boundary, one issue that can arise in  the simulations is the slowing down of autocorrelation times of local observables.
Here we show that the SSE (with directed-loop updates) simulations of the BW-EH do not suffer from this problem.

The efficiency of the SSE scheme in generating independent configurations is probed by the autocorrelation time. For a quantity $Q$, the normalized autocorrelation function is defined as \cite{evertz2003,landaubinder2005}
\begin{equation}
 A(t) = \frac{\left<Q(i+t)Q(i)\right> - \left<Q(i)\right>^2}{\left<Q(i)^2\right> - \left<Q(i)\right>^2},
\end{equation}
where $\left<Q(i)\right>$ and $\left<Q(i+t)Q(i)\right>$ are  averaged values of $Q$ performed in two different MC steps separated by $t$.
The definition of the unit of MC step used here is the same employed in Ref.~\onlinecite{sandvik2002}.
Asymptotically, the autocorrelation function decays exponentially  $e^{-t/\tau}$, where $\tau$ is the the autocorrelation time.
Measurements of $Q$ is independent when $t$ exceeds $\tau$.
Furthermore, the integrated autocorrelation time is defined as
\begin{equation}
 \tau_{int} = \frac{1}{2}  + \sum_{t=1}^{\infty} A(t).
\end{equation}

\begin{figure}[t]
\includegraphics[width=0.85\columnwidth]{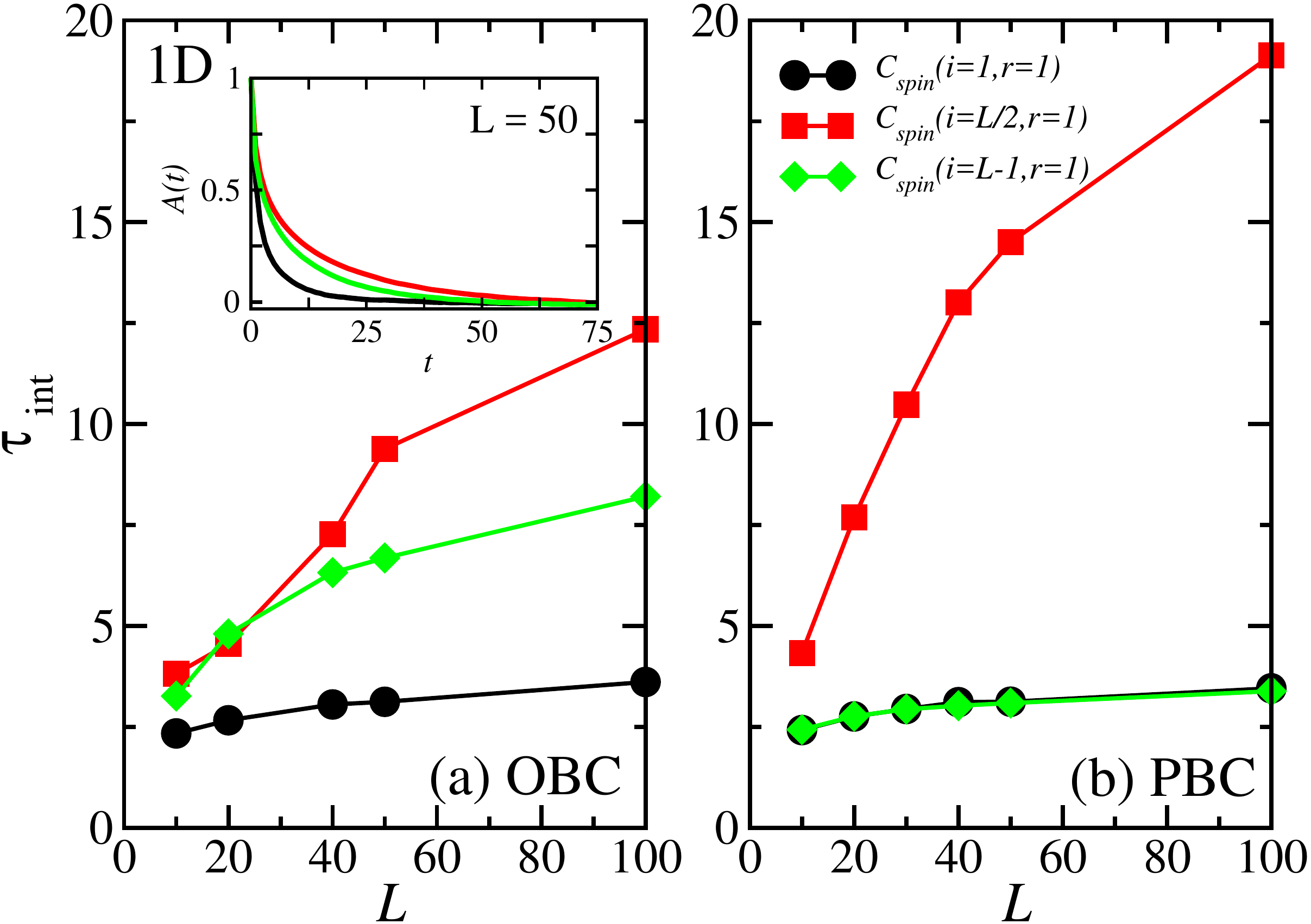}
\centering
\caption{Integrated autocorrelation time, $\tau_{int}$, of $C_{spin}(i,r=1)$ as function of system size, $L$ for (a) OBC (BW) and (b) PBC (CFT1).
The different points, circle (black curve), square (red curve), and diamond (green curve) respectively represent positions $i=1$, $i=L/2$, and $i=L-1$ in the chain.
The inset in panel (a) shows $A(t)$ as a function of $t$ for $L = 50$ and OBC.}
\label{autocorrelation1d} 
\end{figure}

Fig.~\ref{autocorrelation1d} shows the integrated autocorrelation time of the first-neighbour correlation function, $C(i,r=1)$ of a one-dimensional system
for different positions.
We consider both the OBC and PBC entanglement Hamiltonians.
Near the boundary,  $\tau_{int} \approx 3$, and it is almost independent of system size.
In contrast, for the farthest bonds from the boundary, the increase of $\tau_{int}$ with  $L$ is stronger,
and $\tau_{int} \approx 20$ for the largest system considered here, 
which is related to the increase of the effective local ``beta'' of the EH.
It is important to mention, however, that the typical number of MC measurements used here is $N_{meas} \gg \tau_{int}$ ($N_{meas} \approx 10^{8}$).
Thus we can obtain very precise estimators for $C_{spin}(i,r=1)$.

\begin{figure}[b]
\includegraphics[width=0.85\columnwidth]{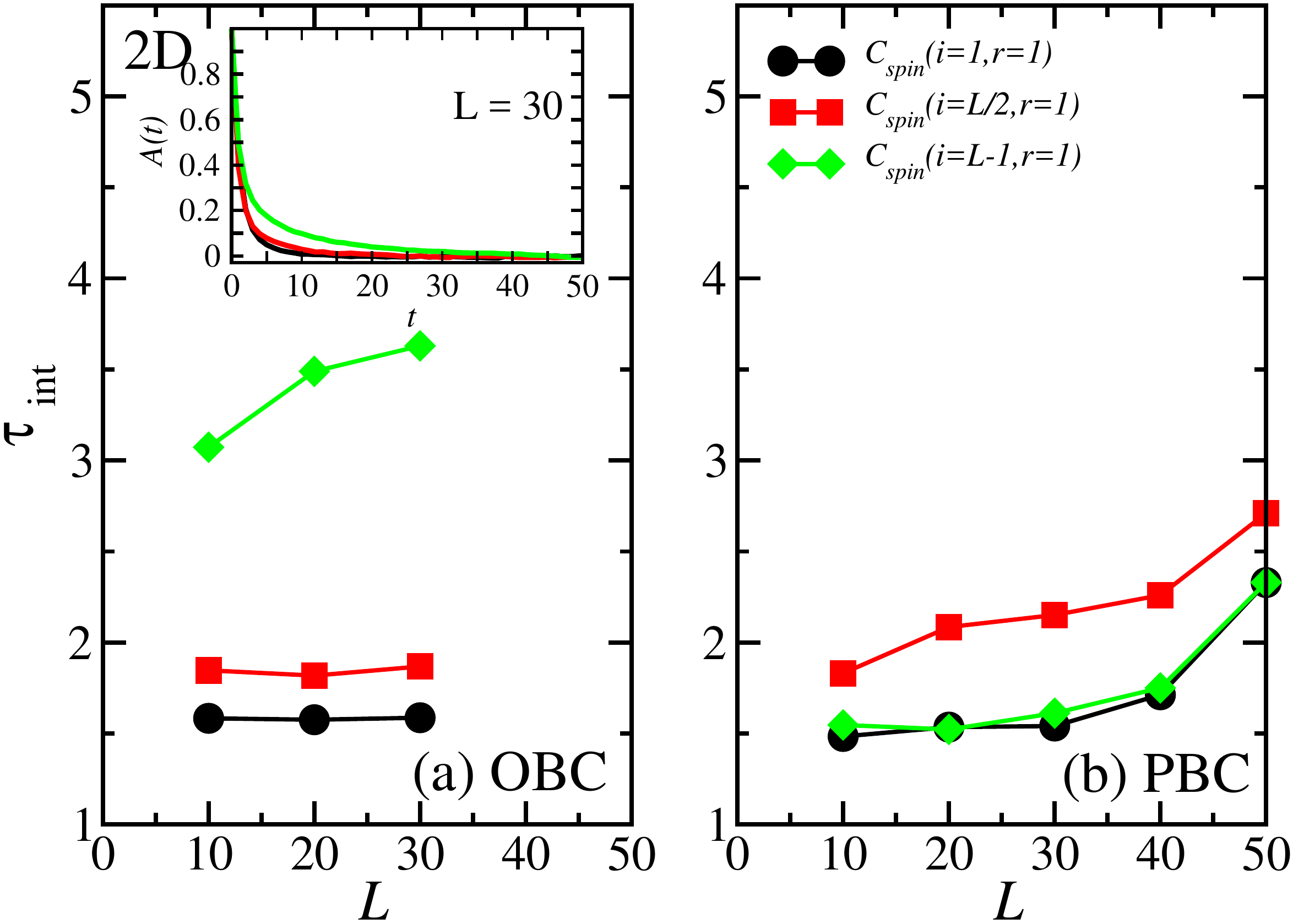}
\centering

\caption{Same plot of Fig. \ref{autocorrelation1d} for a two-dimensional system.
The different points, circle (black curve), square (red curve), and diamond (green curve) respectively represent positions $(i=1,1)$, $(i=L/2,1)$, and $(i=L-1,1)$ in the  square lattice.
The inset in panel (a) shows $A(t)$ as a function of $t$ for $L = 30$ and OBC.}
\label{autocorrelation2d} 
\end{figure}

Similarly to the 1D case, the autocorrelation times of the first-neighbour correlation in 2D is 
much smaller then the typical number of MC measurements considered, $N_{meas} \gg \tau_{int}$, see Fig. \ref{autocorrelation2d}.
In this case, $\tau_{int}$ is even smaller then the values obtained for  1D systems.
This difference between the 1D and the 2D simulations is related to the the loop sizes
built in the directed-loop updates.
In two dimensions the loops are much larger \cite{sandvik2002},  improving the efficiency of the 
simulation of the BW-EH.

\end{document}